\documentclass[a4paper,UKenglish]{lipicsnew}

\def\<<{\ouvreguillemet\everypar={\ouvreguillemet\ }}
\def\ouvreguillemet{\leavevmode\hbox{(\kern-0.20em(\kern+0.20em}\nobreak}
\def\>>{\fermeguillemet\everypar={}}
\def\fermeguillemet{\nobreak\leavevmode\hbox{\kern+0.20em)\kern-0.20em)}}

\newcommand{\bas}[1]{\mbox{\raisebox{-0.5ex}{\scriptsize \it #1}}}
\newcommand{\mbas}[1]{\mbox{\raisebox{-0.75ex}{\scriptsize \it #1}}}

\newcommand{\Bas}[1]{\mbox{\raisebox{-1ex}{\tiny \rm #1}}}
\newcommand{\arrow}{\mbox{\small$\longrightarrow$}}
\newcommand{\BigArrow}{\mbox{\small$-\!\!\!-\!\!\triangleright$}}
\newcommand{\DoubleBigArrow}{\mbox{\small$\triangleleft\!\!-\!\!\!\!\!-\!\triangleright$}}

\newcommand{\deriv}{\mbox{\small$\longleftrightarrow$}}

\newcommand{\bouboule}{\mbox{\raisebox{0.175ex}{\tiny$\,\bullet\,$}}}

\newcommand{\fleche}[1]{\mbox{$\mathop{\arrow}\limits^{#1}$}}

\newcommand{\equiva}{\mbox{$\sim$}}
\newcommand{\smallequiva}{\mbox{\footnotesize$\sim$}}
\newcommand{\smallapprox}{\mbox{\footnotesize$\approx$}}
\newcommand{\BigFleche}[1]{\mbox{$\mathop{\BigArrow}\limits^{#1}$}}
\newcommand{\DoubleBigFleche}[1]{\mbox{$\mathop{\DoubleBigArrow}\limits^{#1}$}}

\newcommand{\gene}[2]{\mbox{$\mathop{\deriv}\limits^{#1}
                                                       _{\mbox{\tiny$#2$}}$}}
\newcommand{\arrowtriangl}{\mbox{$\mathrel=\kern-6pt\rhd$}}

\newcommand{\enonce}[2]{\par{\leftskip0em\vspace{1em}\noindent
                             {\bf #1\ }{\it #2}\par\vspace{1em}}
                        \noindent{\bf Proof.}\\[0.0em]}
\newcommand{\diese}{\mbox{\footnotesize \#}}
\newcommand{\smalldiese}{\mbox{\tiny \#}}

\newcommand{\croix}{\mbox{\scriptsize $\times$}}

\newcommand{\inter}[1]{\hbox{{\rm [}\hskip -2pt {\rm [}{$#1$}{\rm ]}\hskip -2pt {\rm ]}}}

\newcommand{\interFootnote}[1]{\footnotesize\hbox{{\rm [}\hskip -1.5pt
{\rm [}{$#1$}{\rm ]}\hskip -1.5pt {\rm ]}}}

\newcommand{\compose}{\mbox{\tiny o}}

\title{Structural characterization of Cayley graphs}
\titlerunning{A characterization of Cayley graphs}
\author[1]{Didier Caucal}
\affil[1]{CNRS, LIGM, University Paris-Est, France\\
  \texttt{didier.caucal@univ-mlv.fr}}
\authorrunning{D. Caucal}

\begin{document}

\maketitle

\begin{abstract}
{\noindent}We show that the directed labelled Cayley graphs coincide with the 
rooted deterministic vertex-transitive simple graphs. The Cayley graphs are 
also the strongly connected deterministic simple graphs of which all vertices 
have the same cycle language, or just the same elementary cycle language. 
Under the assumption of the axiom of choice, we characterize the Cayley graphs 
for all group subsets as the deterministic, co-deterministic, 
vertex-transitive simple graphs.
\end{abstract}

\section{Introduction}

A group is a basic algebraic structure that comes from the study of polynomial
equations by Galois in 1830. 
To describe the structure of a group, Cayley introduced in 1878 \cite{Ca} 
the concept of graph for any group \,$\mathsf{G}$ \,according to any generating 
subset \,$\mathsf{H}$. 
This is simply the set of labelled oriented edges \,$g\ \fleche{h}\ gh$ \,for
every \,$g$ \,of \,$\mathsf{G}$ \,and \,$h$ \,of \,$\mathsf{H}$. 
Such a graph, called Cayley graph (or Cayley diagram), is directed and 
labelled in \,$\mathsf{H}$ \,(or an encoding of \,$\mathsf{H}$ \,by symbols 
called letters or colors). 
A characterization of unlabelled and undirected Cayley graphs was given by
Sabidussi in 1958 \cite{Sa}. These are the connected graphs whose 
automorphism group has a subgroup with a free and transitive action on the
graph. 
So if one wants to know whether an unlabelled and undirected graph is a Cayley
graph, we must know if we can extract a subgroup of the automorphism group
that allows to define a free and transitive action on the graph. 
To better understand the structure of Cayley graphs, it is pertinent to look 
for characterizations by simple graph-theoretic conditions. 
This approach was clearly stated by Hamkins in 2010: Which graphs are
Cayley graphs?\\
Every Cayley graph is a graph with high symmetry: it is vertex-transitive
meaning that the action of its automorphism group is transitive, or 
equivalently that all its vertices are isomorphic meaning that we `see' the 
same structure regardless of the vertex where we `look'. 
We can characterize the Cayley graphs as vertex-transitive graphs. 
By definition, any Cayley graph satisfies three basic graph properties. 
First is is simple: there are no two arcs of the same source and goal. 
It is also deterministic: there are no two arcs of the same source and label. 
Finally the identity element is a root. 
In this article, we show that these three conditions added to the condition of
being vertex-transitive characterize exactly the Cayley graphs. 
This improves Sabidussi's characterization that easily can be adapted to 
directed and labelled graphs: the Cayley graphs are the deterministic rooted 
simple graphs whose automorphism group has a subgroup with a free and 
transitive action on the graph. In other words, we reduce this last condition 
to the fact that the graph is vertex-transitive. 
For such a simplification, the key result is that every strongly connected,
deterministic and co-deterministic graph is isomorphic to the canonical graph 
of any of its cycle languages. 
This is a fairly standard result in automata theory.\\
Precisely an automaton is just a directed labelled graph (finite or not) with
input and output vertices. 
It recognizes the language of the labels of paths from an input to an output
vertex. Any language \,$L$ \,is recognized by its canonical automaton, namely
the automaton whose graph is the set of transitions between the (left) 
residuals of \,$L$, having \,$L$ \,as its unique initial vertex, and the final
vertices are the residuals of \,$L$ \,containing the empty word. 
We minimize an automaton by identifying its bisimilar vertices. 
Any minimal deterministic and reduced automaton is isomorphic to the 
canonical automaton of its recognized language. 
Moreover, any co-deterministic and reduced automaton is minimal. 
The previous key result follows from these last two properties: any
deterministic and co-deterministic reduced automaton is isomorphic to the 
canonical automaton of its recognized language.\\
An equivalent characterization of Cayley graphs is obtained by strengthening 
the condition of being rooted by the strong connectivity, and simplify the
condition of being vertex-transitive by the fact that all vertices have the
same elementary cycle language. 
Finally, we consider the extension of Cayley graphs for all non-empty subsets
of groups. Under the assumption of the axiom of choice, we show that these
graphs are exactly the deterministic, co-deterministic, vertex-transitive
simple graphs.

\section{Automata}

An automaton is just a directed labelled graph with input and output vertices. 
An accepting path is a path from an initial vertex to a final vertex. 
An automaton recognizes the language of accepting path labels. We recall the
notions of minimal automaton of an automaton, and the notion of canonical
automaton of a language. For any deterministic and reduced automaton, its
minimal automaton is isomorphic to the canonical automaton of its recognized
language. Moreover, any co-deterministic and co-accessible automaton is minimal.
It follows that every strongly connected deterministic and co-deterministic 
graph is isomorphic to the canonical graph of the path language between any two
vertices.

\subsection{Definitions}

We recall basic definitions for directed labelled graphs and automata.\\[0.5em]
Let \,$A$ \,be an arbitrary (finite or infinite) set of symbols. 
We consider a graph as a set of directed edges labelled in \,$A$. 
A directed \,$A$-{\it graph} \,$(V,G)$ \,is defined by a set \,$V$ \,of 
{\it vertices} \,and a subset \,$G \,\subseteq \,V{\croix}A{\croix}V$ \,of 
{\it edges}. 
Any edge \,$(s,a,t) \in G$ \,is from the {\it source} \,$s$ \,to the 
{\it goal} \,$t$ \,with {\it label} \,$a$, and is also written by the 
{\it transition} \,$s\ \fleche{a}_G\ t$ \,or directly \,$s\ \fleche{a}\ t$ \,if 
\,$G$ \,is clear from the context. 
The sources and goals of edges form the set \,$V_G$ \,of 
{\it non-isolated vertices} \,of \,$G$ \,and we denote by \,$A_G$ \,the set of 
edge labels:\\[0.25em]
\hspace*{2em}$V_G\ =\ \{\ s\ |\ \exists\ a,t\ \ (s\ \fleche{a}\ t \,\vee 
\,t\ \fleche{a}\ s)\ \}$ \ \ \ and \ \ \ 
$A_G\ =\ \{\ a\ |\ \exists\ s,t\ \ (s\ \fleche{a}\ t)\ \}$.\\[0.25em]
Thus \,$V - V_G$ \,is the set of {\it isolated vertices}. 
From now on, we assume that any graph \,$(V,G)$ \,is without isolated vertex 
\,({\it i.e.} $V = V_G$) hence the graph can be identified with its edge 
set \,$G$. We also exclude the empty graph \,$\emptyset$. 
Thus, every graph is a non-empty set of labelled edges. 
As any graph \,$G$ \,is a set, there are no two edges with the same source,
goal and label. We say that a graph is {\it simple} \,if there are no two edges 
with the same source and goal: 
\,$(s\ \fleche{a}\ t \,\wedge \,s\ \fleche{b}\ t)\ \Longrightarrow\ a=b$. 
We denote by \,$G^{-1}\ =\ \{\ (t,a,s)\ |\ (s,a,t) \in G\ \}$ \,the 
{\it inverse} \,of a graph \,$G$. 
A graph is {\it deterministic} \,if there are no two edges with the same source 
and label: \,$(r\ \fleche{a}\ s \,\wedge \,r\ \fleche{a}\ t)\ \Longrightarrow\
s=t$. A graph is {\it co-deterministic} \,if its inverse is deterministic:
there are no two edges with the same goal and label. 
For instance, the graph
\,${\rm Even} \ = \ \{(p,a,q)\ ,\ (p,b,p)\ ,\ (q,a,p)\ ,\ (q,b,q)\}$
\,represented as follows:
\begin{center}
\includegraphics{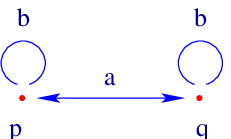}
\end{center}
is deterministic and co-deterministic.
The {\it successor relation} \,$\fleche{}_G$ \,is the unlabelled edge
\,{\it i.e.} \,$s\ \fleche{}_G\ t$ \,if \,$s\ \fleche{a}_G\ t$ \,for some 
\,$a \in A$. 
The {\it accessibility} \,relation \,$\fleche{}^*_G$ \,is the reflexive and 
transitive closure under composition of \,$\fleche{}_G$\,. 
A graph \,$G$ \,is {\it accessible} \,from \,$P \subseteq V_G$ \,if for any
\,$s \in V_G$\,, there is \,$r \in P$ \,such that \,$r\ \fleche{}^*_G\ s$. 
A {\it root} \,is a vertex from which \,$G$ \,is accessible. 
A graph \,$G$ \,is {\it co-accessible} \,from \,$P \subseteq V_G$ \,if 
\,$G^{-1}$ \,is accessible from \,$P$. 
A graph \,$G$ \,is {\it connected} \,if every vertex of \,$G \,\cup \,G^{-1}$ 
\,is a root: \,$s\ \fleche{}^*_{G \,\cup \,G^{-1}}\ t$ \ \ for all 
\,$s,t \in V_G$\,.

Recall that a {\it word} \,$u \,= \,(a_1,\ldots,a_n)$ \,of {\it length}
\,$n \geq 0$ \,over \,$A$ \,is a \,$n$-tuple of letters and denoted for
simplicity \,$a_1{\ldots}a_n$ \,{\it i.e.} \,$u$ \,is a mapping from
\,$\{1,\ldots,n\}$ \,into \,$A$ \,associating with each {\it position}
\,$1 \leq p \leq n$ \,its \,$p$-th letter \,$u(p)$. 
The length of a word \,$u$ \,is denoted by \,$|u|$ \,and for each label \,$a$,
we denote \,$|u|_a \,= \,|\{\ p\ |\ 1 \leq p \leq |u|$ \,and \,$u(p) = a\ \}|$ 
\,the number of positions or {\it occurrences} of \,$a$ \,in \,$u$. 
The word \,$()$ \,of length \,$0$ \,is the {\it empty word} \,and is denoted 
by \,$\varepsilon$. Let \,$A^*\ =\ \{\ (a_1,\ldots,a_n)\ |\ n \geq 0 \,\wedge 
\,a_1,\ldots,a_n \in A\ \}$ \,be the set of words over \,$A$ \,{\it i.e.}
\,$A^*$ \,is the free monoid generated by \,$A$ \,for the concatenation
operation.\\
A {\it language} \,$L$ \,is a set of words \,{\it i.e.} \,$L \subseteq A^*$ 
\,and 
\,$A_L \ = \ \{\ a \in A\ |\ \exists\ u,v \in A^* \ (uav \in L)\ \}$ 
\,is its alphabet. For any \,$u \in A^*$, the language 
\,$u^{-1}L \,= \{\ v\ |\ uv \in L\ \}$ \,is the left {\it residual} \,of \,$L$
\,by \,$u$. 
For all words \,$u,v \in A^*$, $vu$ \,is a {\it conjugated word} \,of \,$uv$. 

A {\it path} \,$(s_0,a_1,s_1,\ldots,a_n,s_n)$ \,of {\it length} \,$n \geq 0$
\,in a graph is a sequence \,$s_0\ \fleche{a_1}\ s_1 \ldots \fleche{a_n}\ s_n$ 
\,of \,$n$ \,consecutive edges, and we write 
\,$s_0\ \fleche{a_1{\ldots}a_n}\ s_n$ \,for indicating the source \,$s_0$\,,
\,the goal \,$s_n$ \,and the label word \,$a_1{\ldots}a_n$ \,of the path. 
A {\it cycle} \,at a vertex \,$s$ \,is a path of source and goal~\,$s$. 
A graph \,$G$ \,is {\it strongly connected} \,if every vertex is a root: 
\,$s\ \fleche{}^*_G\ t$ \ \ for all \,$s,t \in V_G$\,.\\
The set of words labelling the paths from \,$s$ \,to \,$t$ \,of a graph \,$G$ 
\,is\\[0.25em]
\hspace*{12em}${\rm L}_G(s,t)\ =\ \{\ u\ |\ s\ \fleche{u}_G\ t\ \}$\\[0.25em]
the {\it path language} \,of \,$G$ \,from \,$s$ \,to \,$t$. 
For the previous graph \,Even, the path languages are\\[0.25em]
\hspace*{1em}\begin{tabular}{rclcll}
${\rm L}_{\rm Even}(p,p)$ & $=$ & ${\rm L}_{\rm Even}(q,q)$ & $=$ & 
$\{\ u \in \{a,b\}^*\ |\ |u|_a \,\equiv \,0 \,\ ({\rm mod} \ 2)\ \}$ &
denoted \ ${\rm L}_{\rm Even}$\\[0.25em]
${\rm L}_{\rm Even}(p,q)$ & $=$ & ${\rm L}_{\rm Even}(q,p)$ & $=$ & 
$\{\ u \in \{a,b\}^*\ |\ |u|_a \,\equiv\, 1 \,\ ({\rm mod} \ 2)\ \}$ &
denoted \ ${\rm L}'_{\rm Even}$\,.
\end{tabular}\\[0.25em]
The {\it cycle language} \,at vertex \,$s$ \,is 
\,${\rm L}_G(s,s)\ =\ \{\ u\ |\ s\ \fleche{u}_G\ s\ \}$ \,the set of labels of 
cycles at \,$s$\,; in particular \,$\varepsilon \in {\rm L}_G(s,s)$. 
We say that a (non-empty) graph\\[0.25em]
\hspace*{6em}
$G$ \,is a {\it circular graph} \ if \ ${\rm L}_G(s,s) \,= \,{\rm L}_G(t,t)$ \ 
for all \,$s,t \in V_G$\\[0.25em]
and in that case, we denote by \,${\rm L}_G$ \,this common cycle language. 
In other words, a graph is circular if we read the same cycle labels from any 
vertex. The graph \,Even \,is circular.
Every acyclic graph \,$G$ \,is circular and of language
\,${\rm L}_G = \{\varepsilon\}$.\\
The path relation of a deterministic graph is a residual operation for 
recognized languages.
\begin{lemma}\label{PathRes}
For any \,$s\ \fleche{u}_G\ t$ \,and \,$F \subseteq V_G$ \,with \,$G$ 
\,deterministic, \,${\rm L}_G(t,F) \,= \,u^{-1}{\rm L}_G(s,F)$.
\end{lemma}
An {\it automaton} \,${\cal A} = (G,I,F)$ \,is a graph \,$G$ \,with a subset
\,$I \subseteq V_G$ \,of {\it initial vertices} \,and a subset
\,$F \subseteq V_G$ \,of {\it final vertices}. 
The {\it language recognized} by \,${\cal A}$ \,is\\[0.25em]
\hspace*{2em}${\rm L}({\cal A})\ =\ {\rm L}_G(I,F)\ =\ 
\bigcup_{i \in I,f \in F}{\rm L}_G(i,f)\ =\ 
\{\ u\ |\ \exists\ i \in I\ \exists\ f \in F\ (i\ \fleche{u}_G\ f)\ \}$.
\\[0.25em]
An automaton \,${\cal A} = (G,I,F)$ \,is {\it accessible} \,(resp.
{\it co-accessible}) if \,$G$ \,is accessible from \,$I$ \,(resp.
co-accessible from~\,$F$); \,${\cal A}$ \,is {\it reduced} \,if it is
accessible and co-accessible. 
We say that \,${\cal A}$ \,is {\it deterministic} \,if \,$G$ \,is 
deterministic and \,$|I| = 1$. 
Similarly \,${\cal A}$ \,is {\it co-deterministic} \,if \,$G$ \,is 
co-deterministic and \,$|F| = 1$. 
Two automata \,${\cal A}$ \,and \,${\cal B}$ \,are {\it equivalent} \,if they
recognize the same language: \,${\rm L}({\cal A}) = {\rm L}({\cal B})$.

\subsection{Minimal automata}

We reduce an automaton by identifying bisimilar vertices. 
For any deterministic co-accessible automaton, the bisimulation coincides with 
Nerode's congruence. 
Any co-deterministic and co-accessible automaton is minimal and its
determinization remains minimal.\\[0.5em]
Let us consider automata \,${\cal A} \,= \,(G,I,F)$ \,and 
\,${\cal A}' \,= \,(G',I',F')$.\\[0.25em]
A {\it simulation} \,from \,$\cal A$ \,into \,${\cal A}'$ \,is a relation
\,$R \,\subseteq \,V_G{\croix}V_{G'}$ \,such that
\begin{center}
\begin{tabular}{rcl}
$s \in V_{G}$ & $\Longrightarrow$ & $\exists\ s'\ (s\ R\ s')$\\[0.25em]
$(s\ R\ s' \,\wedge \,s\ \fleche{a}_{G}\ t)$ & $\Longrightarrow$ &
$\exists\ t'\ (s'\ \fleche{a}_{G'}\ t' \,\wedge \,t\ R\ t')$\\[0.25em]
$s \in I$ & $\Longrightarrow$ & $\exists\ s'\ (s\ R\ s' \,\wedge \,s' \in I')$
\\[0.25em]
$(s\ R\ s' \,\wedge \,s \in F)$ & $\Longrightarrow$ & $s' \in F'$
\end{tabular}
\end{center}
we say that \,${\cal A}$ \,is {\it simulated by} \,${\cal A}'$ \,and then any
word recognized by \,${\cal A}$ \,is recognized by \,${\cal A}'$\,:
\begin{center}
if \ ${\cal A}$ \,is simulated by \,${\cal A}'$ \ then \
${\rm L}({\cal A}) \,\subseteq \,{\rm L}({\cal A}')$.
\end{center}
A {\it morphism} \,$h$ \,from \,${\cal A}$ \,into \,${\cal A}'$ \,is a mapping 
from \,$V_{G}$ \,into \,$V_{G'}$ \,which is a simulation:
\begin{center}
$s\ \fleche{a}_{G}\ t \ \ \Longrightarrow \ \ h(s)\ \fleche{a}_{G'}\ h(t)$ \ \ 
and \ \ $s \in I \ \Longrightarrow \ h(s) \in I'$ \ \ and \ \ 
$s \in F \ \Longrightarrow \ h(s) \in F'$.
\end{center}
A simulation from \,${\cal A}$ \,into \,${\cal A}'$ \,whose the inverse 
relation is also a simulation is a {\it bisimulation} from \,${\cal A}$ \,on
\,${\cal A}'$ \,and we say that \,${\cal A}$ \,and \,${\cal A}'$ \,are 
{\it bisimilar}\,; \,in this case, they recognize the same language. 
A bisimulation of \,${\cal A}$ \,is a bisimulation from \,${\cal A}$ \,on 
\,${\cal A}$.\\
A {\it reduction} \,$h$ \,from \,${\cal A}$ \,into \,${\cal A}'$ \,is a mapping 
from \,$V_{G}$ \,into \,$V_{G'}$ \,which is a bisimulation, and we write 
\,${\cal A}\ \BigFleche{}_h\ {\cal A}'$ \,or directly 
\,${\cal A}\ \BigFleche{}\ {\cal A}'$ \,if we do not specify a reduction. 
Thus, a reduction is a morphism whose inverse relation is a bisimulation.\\
Therefore two automata are bisimilar if and only if they are reducible into a
same automaton.\\
An injective reduction \,$h$ \,from \,${\cal A}$ \,into \,${\cal A}'$ \,is an 
{\it isomorphism} \,and we write \,${\cal A}\ \DoubleBigFleche{}_h\ {\cal A}'$ 
\,or directly \,${\cal A}\ \DoubleBigArrow{}\ {\cal A}'$.\\
A {\it congruence} \,of \,${\cal A}$ \,is an equivalence on \,$V_{G}$ \,which 
is a bisimulation of \,${\cal A}$.\\
The {\it quotient} \,of \,${\cal A}$ \,by a congruence \,$\sim$ \,is the 
automaton 
\,$([G]\bas{\smallequiva},[I]\bas{\smallequiva},[F]\bas{\smallequiva})$ \,with
\begin{center}
$[G]\bas{\smallequiva} \ = \ \{\ [s]\bas{\smallequiva}\ \fleche{a}\ 
[t]\bas{\smallequiva}\ |\ s\ \fleche{a}_G\ t\ \}$ \ \ and \ \ 
$[S]\bas{\smallequiva} \,= \,\{\ [s]\bas{\smallequiva}\ |\ s \in S\ \}$ \ for 
any \,$S \subseteq V_G$\,.
\end{center}
which is reductible from \,${\cal A}$\,: 
\,${\cal A}\ \BigFleche{}_{\mbox{\tiny{\rm [}\ \,{\rm ]}}}\ 
[{\cal A}]\bas{\smallequiva}$\,. 
Thus \,${\cal A}$ \,and its quotient \,$[{\cal A}]\bas{\smallequiva}$ \,under 
a congruence \,$\sim$ \,recognize the same language. 
The family \,BiSim$({\cal A})$ \,of bisimulations of \,${\cal A}$ \,is closed
under arbitrary union, inverse and composition. Let
\begin{center}
$\approx_{\cal A}\ \ = \ \bigcup\,{\rm BiSim}({\cal A})$
\end{center}
be the greatest bisimulation of \,${\cal A}$ \,which is also the greatest 
congruence of \,${\cal A}$.\\
The {\it minimal automaton} \,Min$({\cal A})$ \,of \,${\cal A}$ \,is the
quotient of \,${\cal A}$ \,under its greatest bisimulation\,:
\begin{center}
${\rm Min}({\cal A}) \ = \ [{\cal A}]\bas{\smallapprox}$\,.
\end{center}
Therefore two automata are bisimilar if and only if their minimal automata are
isomorphic.\\
An automaton \,${\cal A}$ \,is minimal if \,$\approx_{\cal A}$ \,is the identity
\,{\it i.e.} \,${\rm Min}({\cal A})$ \,is isomorphic to \,${\cal A}$.\\
For \,${\cal A}$ \,deterministic and co-accessible, its greatest bisimulation
is Nerode's congruence \cite{Ne}.
\begin{lemma}\label{Nerode}
For any co-accessible automaton \,${\cal A} = (G,I,F)$ \,with \,$G$
\,deterministic,\\[0.25em]
\hspace*{8em}$s\ \approx_{\cal A}\ t \ \ \ \Longleftrightarrow \ \ \ 
{\rm L}_G(s,F) \,= \,{\rm L}_G(t,F)$ \ \ for all \,$s,t \in V_{\cal A}$\,.
\end{lemma}
For any graph \,$G$, we denote by \,$P{\cdot}u\ =\ \{\ t\ |\ \exists\ s \in P\
(s\ \fleche{u}_G\ t)\ \}$ \,the set of vertices accessible from a vertex in
\,$P \subseteq V_G$ \,by a path in \,$G$ \,labelled by \,$u \in A^*$.\\
We determinize any automaton \,${\cal A} = (G,I,F)$ \,into the following
automaton \,${\rm Det}({\cal A})$\,:
\begin{center}
$\bigl(\{\ I{\cdot}u\ \fleche{a}\ I{\cdot}ua\ |\ u \in A^* \,\wedge \,a \in A 
\,\wedge \,I{\cdot}ua \neq \emptyset\ \}\,,\,\{I\}\,,\,\{\ I{\cdot}u\ |\ 
u \in A^* \,\wedge \,I{\cdot}u \,\cap F \neq \emptyset\ \}\bigr)$
\end{center}
which is deterministic, accessible and recognizes \,${\rm L}({\cal A})$. 
Moreover \,${\rm Det}({\cal A})$ \,is co-accessible when \,${\cal A}$ \,is
co-accessible, and minimal if in addition \,${\cal A}$ \,is co-deterministic.
\begin{lemma}\label{CoDetMin}
For any automaton \,${\cal A}$ \,co-deterministic and co-accessible,\\
\hspace*{6em}${\cal A}$ \,and \,${\rm Det}({\cal A})$ \,are minimal.
\end{lemma}
For any automaton \,${\cal A} = (G,I,F)$, its {\it inverse} 
\,${\cal A}^{-1} = (G^{-1},F,I)$ \,recognizes the mirrors of the words of
\,${\rm L}({\cal A})$. 
We co-determinize \,${\cal A}$ \,into the equivalent automaton 
\,${\rm CoDet}({\cal A}) \,= \,({\rm Det}({\cal A}^{-1}))^{-1}$ \,which is 
co-deterministic and co-accessible. 
Lemma~\ref{CoDetMin} provides a fairly standard transformation of any
automaton into a deterministic minimal equivalent automaton: we apply the
co-determinization followed by the determinization.
\begin{proposition}\label{DetCoDet}
For any automaton \,${\cal A}$, the automaton 
\,${\rm Det}\bigl({\rm CoDet}({\cal A})\bigr)$ \,is minimal,\\
\hspace*{7.5em}deterministic, reduced and recognizes \,${\rm L}({\cal A})$.
\end{proposition}

\subsection{Canonical automata}

For any language \,$L$ \,and up to isomorphism, there is a unique minimal, 
deterministic and reduced automaton recognizing \,$L$. Such an automaton is 
given by the residual graph of \,$L$ \,with the unique initial vertex \,$L$
\,and the final vertices are the residuals of \,$L$ \,containing the empty 
word. Any reduced, deterministic and co-deterministic automaton \,${\cal A}$ 
is isomorphic to the canonical automaton of the language recognized by 
\,${\cal A}$.\\[0.5em]
To every language \,$L$ \,is associated its {\it canonical graph} \,or 
{\it residual graph}:
\begin{center}
$\overrightarrow{L} \ = \ \{\ u^{-1}L\ \fleche{a}\ (ua)^{-1}L\ |\ u \in A^* 
\,\wedge \,a \in A \,\wedge \,(ua)^{-1}L \neq \emptyset \ \}$.
\end{center}
For instance \,$a^{-1}{\rm L}_{\rm Even} \,= \,{\rm L}'_{\rm Even} \,= 
\,b^{-1}{\rm L}'_{\rm Even}$ \ and 
\ $b^{-1}{\rm L}_{\rm Even} \,= \,{\rm L}_{\rm Even} \,= 
\,a^{-1}{\rm L}'_{\rm Even}$\,. 
Thus \,$\overrightarrow{{\rm L}_{\rm Even}} \ = \ 
\overrightarrow{{\rm L}'_{\rm Even}}$ \,is the following graph which is
isomorphic to the graph \,${\rm Even}$\,:
\begin{center}
\includegraphics{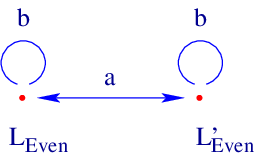}
\end{center}
The {\it canonical automaton} \,of any language \,$L$ \,is the automaton
\begin{center}
${\rm Can}(L)\ =\ 
\bigl(\overrightarrow{L}\,,\,\{L\}\,,\,\{\,u^{-1}L\,|\,u \in L\,\}\bigr)$
\end{center}
which is the unique minimal, deterministic and reduced automaton recognizing
\,$L$.
\begin{lemma}\label{MinDet}
For any deterministic and reduced automaton \,${\cal A}$, the automaton 
\,${\rm Min}({\cal A})$ \,is\\
\hspace*{6em}isomorphic to \,${\rm Can}({\rm L}({\cal A}))$.
\end{lemma}
This Lemma~\ref{MinDet} restricted to finite automata is the Myhill-Nerode 
theorem \cite{Ha,HU}. 
Lemma~\ref{MinDet} with Proposition~\ref{DetCoDet} (or Lemma~\ref{CoDetMin})
imply the isomorphism of equivalent automata which are reduced, deterministic
and co-deterministic.
\begin{proposition}\label{DETCoDET}
For any automaton \,${\cal A}$ \,reduced, deterministic and co-deterministic, 
${\cal A}$ \,is\\
\hspace*{8em}isomorphic to \,${\rm Can}({\rm L}({\cal A}))$.
\end{proposition}
We just see that the graph Even is isomorphic to 
\,$\overrightarrow{{\rm L}_{\rm Even}}$ \,or
\,$\overrightarrow{{\rm L}'_{\rm Even}}$\,. 
This generalizes to any strongly connected, deterministic and co-deterministic
graph \,$G$ \,by applying Proposition~\ref{DETCoDET} to the automaton
\,$(G,s,t)$ \,for every vertices \,$s,t$.
\begin{corollary}\label{EquivRes}
For any graph \,$G$ \,strongly connected, deterministic and co-deterministic,\\
\hspace*{6.5em}$G$ \,is isomorphic to \,$\overrightarrow{{\rm L}_G(s,t)}$
\,for all \,$s,t \in V_G$\,.
\end{corollary}
This corollary implies that any strongly connected, deterministic and 
co-deterministic graph is minimal with respect to any of its cycle languages. 
It follows from Corollary~\ref{EquivRes} that two strongly connected 
deterministic and co-deterministic graphs are isomorphic if they have a same 
path language.
\begin{corollary}\label{EquivGraph}
For any graphs \,$G,H$ \,strongly connected, deterministic and 
co-deterministic,\\
\hspace*{6.5em}if \ \,${\rm L}_G(s,t) \,=\,{\rm L}_H(p,q)$ \,for some 
\,$s,t \in V_G$ \,and \,$p,q \in V_H$ \ then \ $G \,\DoubleBigArrow{} \,H$.
\end{corollary}
%
%
This corollary is a key property to provide a structural characterization of 
Cayley graphs.

\section{Cayley graphs}

Sabidussi's theorem characterizes the undirected and unlabelled Cayley graphs 
as the connected graphs having a free transitive action by a subgroup of the
automorphism group. 
We simply adapt this theorem to directed labelled graphs by replacing the
connectedness with the conditions of being rooted, deterministic and simple.

\subsection{Cayley graphs and Sabidussi's theorem}

Let \,$(\mathsf{G},\cdot)$ \,be a group {\it i.e.} a set \,$\mathsf{G}$ \,with 
an associative internal binary operation \,$\cdot$ \,such that there exists an 
identity element \,$1_{\mathsf{G}}$ \,and each \,$g \in \mathsf{G}$ \,has an
inverse \,$g^{-1}$. 
Let \,$\mathsf{H}$ \,be a non-empty 
generating subset of \,$\mathsf{G}$\,: \ for any \,$g \in \mathsf{G}$, there 
are \,$n \geq 0$ \,and \,$h_1,\ldots,h_n \in \mathsf{H}$ \,such that 
\,$g = h_1{\cdot}{\ldots}{\cdot}h_n$. 
Let \,$\inter{\ }\,: \mathsf{H}\ \fleche{}\ A$ \,be an injective mapping 
coding each \,$h \in \mathsf{H}$ \,by an element \,$\inter{h} \in A$. 
The image of \,$\inter{\ }$ \,is the set 
\,$\inter{\mathsf{H}} \,= \,\{\ \inter{h}\ |\ h \in \mathsf{H}\ \}$ 
\,of labels of \,$\mathsf{H}$.\\
The {\it Cayley graph} \,of \,$(\mathsf{G},\mathsf{H},\inter{\ })$ \,is the 
graph
\begin{center}
${\cal C}\inter{\mathsf{G},\mathsf{H}}\ =\ \{\ g\ \fleche{\interFootnote{h}}\ 
g{\cdot}h\ |\ g \in \mathsf{G} \,\wedge \,h \in \mathsf{H}\ \}$.
\end{center}
This graph is deterministic, co-deterministic, simple and strongly connected:
\\[0.25em]
\hspace*{1.5em}
$g\ \fleche{\interFootnote{h_1}}_{{\cal C}\interFootnote{\mathsf{G},\mathsf{H}}}\ 
g{\cdot}h_1 \ldots
\fleche{\interFootnote{h_n}}_{{\cal C}\interFootnote{\mathsf{G},\mathsf{H}}}\ 
\,g{\cdot}h_1{\cdot}{\ldots}{\cdot}h_n$ \ \ for all 
\,$n \geq 0$ \,and \,$g,h_1,\ldots,h_n~\in~\mathsf{H}$.\\[0.25em]
From this path, we deduce that any Cayley graph is circular and of language
\\[0.25em]
\hspace*{5em}${\rm L}_{{\cal C}\interFootnote{\mathsf{G},\mathsf{H}}} \,= 
\,\{\ \inter{h_1}\ldots\inter{h_n}\ |\ n \geq 0 \,\wedge 
\,h_1,\ldots,h_n \in H \,\wedge \,h_1{\cdot}{\ldots}{\cdot}h_n = 1\ \}$.
\\[0.25em]
By Corollary~\ref{EquivRes}, ${\cal C}\inter{\mathsf{G},\mathsf{H}}$ \,is 
isomorphic to the canonical graph
\,$\overrightarrow{{\rm L}_{{\cal C}\interFootnote{\mathsf{G},\mathsf{H}}}}$\,.\\
A well-known characterization of the unlabelled and non-oriented Cayley graphs 
was given by Sabidussi~\cite{Sa}. Let us recall this characterization.\\
First of all, a left {\it action} \,of \,$\mathsf{G}$ \,on a set \,$V$ \,is a 
mapping \ $\bouboule\ : \,\mathsf{G}{\croix}V\ \fleche{}\ V$ \ associating to
each \,$(g,s) \in \mathsf{G}{\croix}V$ \,the image \,$g{\bouboule}s$ \,such 
that for all \,$s \in V$ \,and \,$g,h \in \mathsf{G}$,\\[0.25em]
\hspace*{12em}$1{\bouboule}s \,= \,s$ \ \ and \ \ 
$h{\bouboule}(g{\bouboule}s) \,= \ (h{\cdot}g){\bouboule}s$\\[0.25em]
Note that for any \,$g \in \mathsf{G}$, the mapping 
\,$\overrightarrow{g}\,: \,s\ \mapsto\ g{\bouboule}s$ \,is a permutation of 
\,$V$. 
Thus, a group action of \,$\mathsf{G}$ \,on \,$V$ \,may be seen as a group 
homomorphism from \,$\mathsf{G}$ \,into the group of permutations of \,$V$. 
We say that the action $\bouboule$ \,is {\it transitive} \,if\\[0.25em]
\hspace*{6em}
for all \,$s,t \in V$, there exists \,$g \in \mathsf{G}$ \,such that 
\,$g{\bouboule}s \,= \,t$.\\[0.25em]
We also say that the action $\bouboule$ \,is {\it free} \,if\\[0.25em]
\hspace*{3em}
for all \,$g,h \in \mathsf{G}$, \ if there exists \,$s \in V$ \,such that 
\,$g{\bouboule}s \,= \,h{\bouboule}s$ \ then \ $g = h$.\\[0.25em]
So a free and transitive action $\bouboule$ \,means that\\[0.25em]
\hspace*{6em}
for all \,$s,t \in V$, there exists a unique \,$g \in \mathsf{G}$ \,such that 
\,$g{\bouboule}s \,= \,t$.\\[0.25em]
Let \,$G$ \,be a (directed and labelled) graph.\\
An action \,of \,$\mathsf{G}$ \,on \,$G$ \,is an action \,$\bouboule$ \,of 
\,$\mathsf{G}$ \,on \,$V_G$ \,which is a morphism of \,$G$ \,{\it i.e.}
\\[0.25em]
\hspace*{3em}
$s\ \fleche{a}_G\ t \ \ \ \Longrightarrow \ \ \ 
g{\bouboule}s\ \fleche{a}_G\ g{\bouboule}t$ \ \ \ for all \,$s,t \in V_G$\,, 
\,$a \in A_G$\,, \,$g \in \mathsf{G}$.\\[0.25em]
Therefore, a group action of \,$\mathsf{G}$ \,on \,$G$ \,may be seen as a group 
homomorphism from \,$\mathsf{G}$ \,into the group \,${\rm Aut}(G)$ \,of 
{\it automorphisms} of \,$G$ \,{\it i.e.} \,of isomorphisms from \,$G$ \,to 
\,$G$.\\
We say that vertices \,$s,t$ \,of a graph \,$G$ \,are {\it isomorphic} \,and
we write \,$s \,\simeq_G t$ \,if there is an automorphism \,$h$ \,of \,$G$
\,such that \,$t = h(s)$.\\
A graph \,$G$ \,is {\it vertex-transitive} \,if there exists a transitive group
action on \,$G$. This means that \,${\rm Aut}(G)$ \,acts transitively on \,$G$
\,or equivalently that all its vertices are isomorphic: $s \,\simeq_G t$ \ for
all \,$s,t \in V_G$\,. In particular, any vertex-transitive graph is circular.\\
First, we adapt to all Cayley graphs a Sabidussi's characterization for a 
given group.
\begin{proposition}\label{CaractCayley3}
A graph \,$G$ \,is isomorphic to a Cayley graph of a group \,$\mathsf{G}$ \,if 
and only if\\
$G$ \,is a deterministic rooted simple graph with a free transitive action of 
\,$\mathsf{G}$ \,on \,$G$.
\end{proposition}
\proof
$\Longrightarrow$\,: \,Assume that 
\,$G \,\DoubleBigArrow{} \,{\cal C}\inter{\mathsf{G},\mathsf{H}}$ \,for some 
generating subset \,$\mathsf{H}$ \,of \,$\mathsf{G}$ \,and some coding 
\,$\inter{\ }$ \,of \,$\mathsf{H}$. 
The vertex set of \,${\cal C}\inter{\mathsf{G},\mathsf{H}}$ \,is 
\,$\mathsf{G}$ \,whose group operation \,$\cdot$ \,is a free transitive action
of \,$\mathsf{G}$ \,on \,${\cal C}\inter{\mathsf{G},\mathsf{H}}$.\\
In particular \,${\cal C}\inter{\mathsf{G},\mathsf{H}}$ \,is vertex-transitive 
for any subset \,$\mathsf{H}$ \,of \,$\mathsf{G}$.\\[0.25em]
$\Longleftarrow$\,: \,Let \,$\bouboule$ \,be a free transitive action of 
\,$\mathsf{G}$ \,on \,$G$.\\
Let us check that \,$G$ \,is isomorphic to a Cayley graph of \,$\mathsf{G}$ 
\,by simply adapting the proof of the sufficient condition of Sabidussi's 
theorem.\\
As \,$G$ \,is rooted, we can pick a root \,$r$ \,of \,$G$.\\
For all \,$s \in V_G$ \,there is a unique \,$g_s \in \mathsf{G}$ \,such that 
\,$g_s{\bouboule}r \,= \,s$.\\
Thus \,$h \,= \,g_{h{\bouboule}r}$ \,for all \,$h \in \mathsf{G}$\,;
\,in particular \,$g_r = 1$. We define\\[0.25em]
\hspace*{12em}$\mathsf{H}\ =\ \{\ g_s\ |\ r\ \fleche{}_G\ s\ \}$.\\[0.25em]
As \,$G$ \,is simple and deterministic, we define the following injection 
\,$\inter{\ }$ \,from \,$\mathsf{H}$ \,into \,$A_G$ \,by\\[0.25em]
\hspace*{12em}
$\inter{g_s} \,= \,a$ \ \ for any \,$r\ \fleche{a}_G\ s$.\\[0.25em]
By renaming each vertex \,$s$ \,of \,$G$ \,by \,$g_s$, we get \,$G$
\,isomorphic to the graph\\[0.25em]
\hspace*{12em}
$\overline{G}\ =\ \{\ g_s\ \fleche{a}\ g_t\ |\ s\ \fleche{a}_G\ t\ \}$.
\\[0.25em]
We check (see Appendix) that
\,$\overline{G} \,= \,{\cal C}\inter{\mathsf{G},\mathsf{H}}$ \,and 
\,$\mathsf{H}$ \,is a generating subset of \,$\mathsf{G}$.
\qed\\[1em]
The determinism condition of this proposition is necessary. 
For instance, the following simple and strongly connected graph \,$G$\,:
\begin{center}
\includegraphics{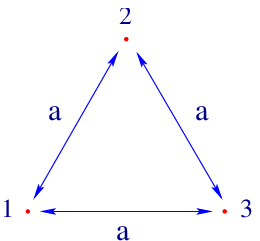}
\end{center}
is not deterministic hence is not a Cayley graph, while this graph has a free 
transitive action of the (cyclic) group of order \,$3$.\\
Proposition~\ref{CaractCayley3} is a restricted characterization of Cayley 
graphs since it is relative to a group~$\mathsf{G}$. 
We say that a graph \,$G$ \,{\it has a free transitive action} \,if there 
exists a group \,$\mathsf{G}$ \,with a free transitive action \,$\bouboule$ 
\,on \,$G$\,; \,in that case 
\,$\{\ \overrightarrow{g}\ |\ g \in \mathsf{G}\ \}$ \,is a subgroup of 
\,${\rm Aut}(G)$ \,and its canonical action 
\,$(\overrightarrow{g},s)\ \mapsto\ \overrightarrow{g}(s) \,= 
\,g \,\bouboule \,s$ \,is free and transitive on \,$G$. 
Proposition~\ref{CaractCayley3} give a simple generalization of 
Sabidussi'theorem to labelled directed graphs.
\begin{proposition}\label{CaractCayley2}
A graph \,$G$ \,is a Cayley graph if and only if \,$G$ \,is deterministic, 
rooted, simple, with a free transitive action.
\end{proposition}
Proposition~\ref{CaractCayley2} characterizes the Cayley graphs using two 
conditions of different nature. The first condition is structural: the graph 
must be deterministic, simple and rooted. The second condition is algebraic 
namely the existence of a free transitive action. 
We now give a characterization that is only structural by restricting the 
algebraic condition to the vertex-transitivity: we no longer need to extract a 
subgroup of the automorphism group whose the canonical action is free and
transitive.

\subsection{Cayley graphs of languages}

We briefly recall the definition of a group by a language whose letters form 
the set of generators and the words define the set of relators \cite{MKS}. 
Let \,$L \subseteq A^*$ \,be a language.\\
The word operation of {\it deleting} \,a word of \,$L$ \,is the rewriting 
according to \,$L{\croix}\{\varepsilon\}$\,:\\[0.25em]
\hspace*{10em}
$xuy\ \fleche{}_L\ xy$ \ \ for any \,$u \in L$ \,and \,$x,y \in A^*$\\[0.25em]
and the inverse operation \,$(\fleche{}_L)^{-1}$ \,is the insertion of a word
of \,$L$.\\
The {\it derivation} \,$\fleche{*}_L$ \,and the {\it Thue congruence} 
\,$\gene{*}{}$$_L$ \,of \,$L$ \,are the reflexive and transitive closure 
under composition of respectively \,$\fleche{}_L$ \,and 
\,$\gene{}{}$$_L \ = \ \fleche{}_L \,\cup \,(\fleche{}_L)^{-1}$\,.\\
The equivalence class of \,$u \in A^*$ \,with respect to 
\,$\gene{*}{}$$_L$ \,is denoted 
\,$[u]$\mbas{$L$} \ $= \ \{\ v\ |\ u\ \gene{*}{}$$_L\ v\ \}$.\\
We say that a language \,$L$ \,is a {\it group presentation language} \,if
\\[0.25em]
\hspace*{6em}\begin{tabular}{ll}
$(i)$ & $A_L \neq \emptyset$ \,{\it i.e.} \,$L \neq \emptyset$ \,and
\,$L \neq \{\varepsilon\}$\\[0.25em]
$(ii)$ & for all \,$a \in A_L$\,, \,there exists $u \in A_L^*$ \,such that 
\,$au,ua \in [\varepsilon]$\mbas{$L$}\\[0.25em]
$(iii)$ & for all \,$a \neq b \in A_L\,,
\,[a]$\mbas{$L$}$ \neq [b]$\mbas{$L$}\,.
\end{tabular}\\[0.25em]
In that case, the quotient \,$\{\ [u]$\mbas{$L$}$\ |\ u \in A_L^*\ \}$ \,of 
\,$A_L^*$ \,under the congruence \,$\gene{*}{}$$_L$ \,is by \,$(ii)$ \,a group 
for the operation\\[0.25em]
\hspace*{12em}
$[u]$\mbas{$L$}${\cdot}[v]$\mbas{$L$}$ \,= \,[uv]$\mbas{$L$} \ for all 
\,$u,v \in A_L^*$\\[0.25em]
and we define \,${\cal C}(L)$ \,as being the Cayley graph of 
\,$\{\ [u]$\mbas{$L$}$\ |\ u \in A_L^*\ \}$ \,generated by the subset 
\,$\{\ [a]$\mbas{$L$}$\ |\ a \in A_L\ \}$\,:
\begin{center}
${\cal C}(L) \ \,= \ 
\,{\cal C}\,\inter{\,\{\,[u]$\mbas{$L$}$\,|\,u \in A_L^*\,\}\,,\,
\{\,[a]$\mbas{$L$}$\,|\,a \in A_L\,\}\,}$
\end{center}
where \,$[a]$\mbas{$L$} \,is encoded by \,$\inter{[a]\mbas{$L$}} = a$ \,for
all \,$a \in A_L$\,; this makes sense from \,$(iii)$ \,and this graph is
non-empty by $(i)$.\\
For instance \,${\cal C}(\{a^6,b^2,(ab)^3\})$ \,is represented by
the following tiling plane:
\begin{center}
\includegraphics{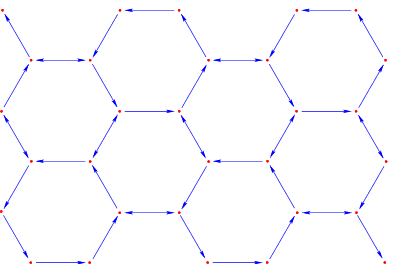}
\end{center}
where every simple (resp. double) arrow is labelled by \,$a$ (resp. $b$), 
and \,${\cal C}(\{a^6,b^3,aba\})$ \,is
\begin{center}
\includegraphics{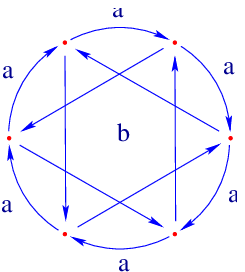}
\end{center}
As \,${\cal C}(L)$ \,is a Cayley graph, it is circular and its language is
\,$[\varepsilon]$\mbas{$L$}\,.
\begin{lemma}\label{LangCayley}
For any group presentation language \,$L$, 
\,${\rm L}_{{\cal C}(L)} \,= \,[\varepsilon]$\mbas{$L$} \,and
\,${\cal C}(L) \,= \,{\cal C}([\varepsilon]$\mbas{$L$}$)$.
\end{lemma}
We say that \,$L$ \,is a {\it stable language} \,if\\[0.25em]
\hspace*{12em}
$uvw \in L \ \Longleftrightarrow \ uw \in L$ \ \ for any \,$v \in L$\\[0.25em]
meaning that \,$L$ \,is preserved by insertion and deletion of factor in \,$L$. 
By iterating these two word operations from \,$\varepsilon$, it only gets all 
words of \,$L$.
\begin{lemma}\label{LangStable}
A non-empty language \,$L$ \,is stable \,if and only if 
\,$L \,= \,[\varepsilon]$\mbas{$L$}\,.
\end{lemma}

\section{Graph characterizations of Cayley graphs}

We begin with basic graph properties, especially for circular graphs. 
We then give a first characterization of Cayley graphs: they are the 
vertex-transitive and rooted deterministic simple graphs (Theorem~\ref{Main}). 
They are also the circular and strongly connected deterministic simple graphs
(Theorem~\ref{MainBis}). We can also replace the circularity by the elementary
circularity because every vertex-transitive graph is elementary circular which
is then circular (Lemma~\ref{CircEleCirc}). 
Another significant characterization concerns Cayley graphs for all subsets of 
groups: under ZFC, they are the deterministic,
co-deterministic, vertex-transitive simple graphs (Theorem~\ref{MainFour}).

\subsection{A first graph characterization}

We consider the family \,${\cal F}$ \,of deterministic, rooted, simple graphs 
which are vertex-transitive. We want to establish that these graphs are Cayley 
graphs. In particular, any graph of \,${\cal F}$ \,should be strongly 
connected.
\begin{lemma}\label{RootTransitive}
Any rooted vertex-transitive graph is strongly connected.
\end{lemma}
Furthermore any graph of \,${\cal F}$ \,should be co-deterministic.
\begin{lemma}\label{CoDet}
Any deterministic and strongly connected circular graph is co-deterministic.
\end{lemma}
By Corollary~\ref{EquivRes}, Lemmas~\ref{RootTransitive} and \ref{CoDet}, 
any graph \,$G$ \,of \,${\cal F}$ \,is isomorphic to the canonical graph of its 
path languages, hence in particular to \,$\overrightarrow{{\rm L}_G}$. 
This cycle language \,${\rm L}_G$ \,is stable.
\begin{lemma}\label{ClasseVide}
For any deterministic circular graph \,$G$, \,${\rm L}_G$ \,is a stable 
language.
\end{lemma}
For any circular graph, the cycle language is closed under conjugacy, and any 
label of the graph is a letter of this language when the graph is strongly 
connected.
\begin{lemma}\label{SymFort}
For any strongly connected circular graph \,$G$,\\[0.25em]
\hspace*{7em}${\rm L}_G$ \,is closed under conjugacy and of letter set 
\,$A_{{\rm L}_G} \,= \,A_G$\,.
\end{lemma}
Let us give a condition on a circular graph for its cycles to form a group
presentation language.
\begin{lemma}\label{CircPres}
For any strongly connected, deterministic circular simple graph \,$G$,\\[0.25em]
\hspace*{7em}${\rm L}_G$ \,is a group presentation language.
\end{lemma}
Any vertex-transitive graph is circular. 
By Lemma~\ref{CoDet} and Corollary~\ref{EquivRes}, the converse is true when 
the graph is deterministic and strongly connected.
\begin{lemma}\label{CircSym}
For any deterministic and strongly connected graph \,$G$,\\
\hspace*{7em}$G$ \,is vertex-transitive if and only if \ $G$ \,is circular.
\end{lemma}
We are able to establish a first structural characterization of Cayley graphs.
\begin{theorem}\label{Main}
A graph is a Cayley graph if and only if it is deterministic, rooted, simple\\
\hspace*{7.5em}and vertex-transitive.
\end{theorem}
\proof
$\Longrightarrow$\,: \,By definition, a Cayley graph is strongly connected, 
deterministic, simple and circular. 
As already indicated in the proof of Proposition~\ref{CaractCayley3} (or by
Lemma~\ref{CircSym}), $G$ \,is vertex-transitive.\\[0.25em]
$\Longleftarrow$\,: \,Let \,$G$ \,be a deterministic, rooted, simple and
vertex-transitive graph.\\
By Lemma~\ref{RootTransitive}, $G$ \,is strongly connected. 
By Lemma~\ref{CoDet}, $G$ \,is co-deterministic.\\
By Lemma~\ref{CircPres}, \,${\rm L}_G$ \,is a group presentation hence
\,${\cal C}({\rm L}_G)$ \,is a Cayley graph.\\
By Lemma~\ref{LangCayley}, we have \,${\rm L}_{{\cal C}({\rm L}_G)} \,=
\,[\varepsilon]$\mbas{${\rm L}_G$}\,.\\
By Lemmas~\ref{ClasseVide} and \ref{LangStable},
\,${\rm L}_G \,= \,[\varepsilon]$\mbas{${\rm L}_G$}\,.\\
By Corollary~\ref{EquivGraph}, \,$G$ \,is isomorphic to 
\,${\cal C}({\rm L}_G)$ \,hence \,$G$ \,is a Cayley graph.
\qed\\[1em]
A deterministic graph of Petersen skeleton is given by the following two 
isomorphic representations:
\begin{center}
\includegraphics{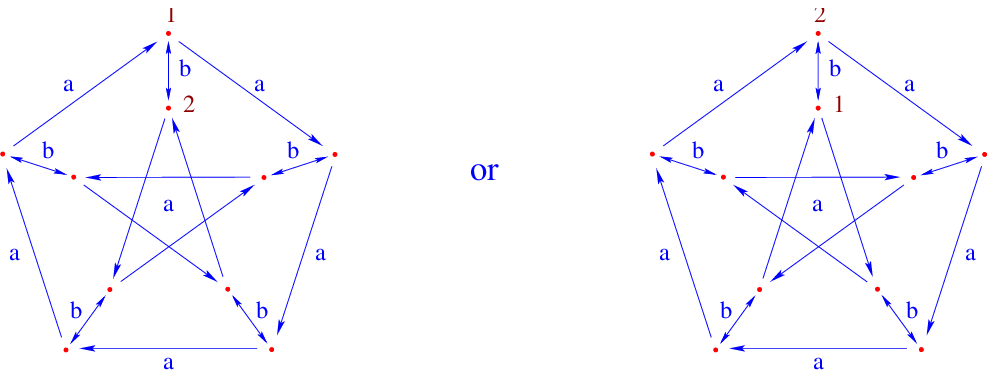}
\end{center}
Such a graph is not a Cayley graph because it is not circular (hence not 
vertex-transitive):
\begin{center}
$ababa \in {\rm L}_{\rm Petersen}(1,1) - {\rm L}_{\rm Petersen}(2,2)$ \ \ and \ \ 
$baaba \in {\rm L}_{\rm Petersen}(2,2) - {\rm L}_{\rm Petersen}(1,1)$.
\end{center}
Other condition such as the same in and out degrees is a consequence of 
Theorem~\ref{Main}. 
Precisely a graph \,$G$ \,is source-{\it complete} \,if for all vertex \,$s$
\,and label \,$a$, there is an edge from \,$s$ \,labelled by \,$a$\,:\\
\hspace*{12em}
$\forall\ s \in V_G\ \ \forall\ a \in A_G\ \ \exists\ t \ (s\ \fleche{a}\ t)$.
\\[0.25em]
A graph is {\it co-complete} \,if its inverse is complete.\\
Let us give weaker conditions than those of Theorem~\ref{Main} to
obtain a complete graph.
\begin{lemma}\label{Complete}
Any strongly connected circular graph is complete and co-complete.
\end{lemma}
Note also that we can replace the vertex-transitivity in
Theorem~\ref{Main} with edge-transitivity.
Recall that a graph \,$G$ \,is {\it edge-transitive} \,if for all
\,$(s,a,t)\,,\,(s',a,t') \in G$, there exists an automorphism \,$h$ \,of \,$G$
\,such that \,$h(s) = s'$ \,and \,$h(t) = t'$. 
The determinism with the completeness give the correspondence between the 
vertex-transitivity and the edge-transitivity.
\begin{lemma}\label{EdgeTrans}
For any graph \,$G$ \,deterministic and complete,\\
\hspace*{7em}$G$ \,is vertex-transitive \ \ $\Longleftrightarrow$ \ \ $G$ \,is 
edge-transitive.
\end{lemma}

\subsection{Other graph characterizations}

Let us give another characterization of Cayley graphs by replacing the
vertex-transitivity condition by the circularity or the elementary circularity
which is now defined.\\
The {\it elementary cycle language} at vertex \,$s$ \,is the label set\\[0.25em]
\hspace*{1em}
${\rm E}_G(s)\ =\ \{\ a_1{\ldots}a_n\ |\ n > 0 \,\wedge \,\exists\ 
s_0 \neq \ldots \neq s_{n-1}\ (s_0\ \fleche{a_1}\ s_1 \ldots \fleche{a_{n-1}}\ 
s_{n-1}\ \fleche{a_n}\ s_0 = s)\ \}$\\[0.25em]
of elementary cycles (passing only by distinct vertices) at \,$s$\,; in 
particular \,$\varepsilon \not\in {\rm E}_G(s) \subset {\rm L}_G(s,s)$. 
We say that the (non-empty) graph \,$G$ \,is an
{\it elementary circular graph} \,if\\[0.25em]
\hspace*{10em}
${\rm E}_G(s) \,= \,{\rm E}_G(t)$ \ for all \,$s,t \in V_G$\\[0.25em]
and in that case, we denote by \,${\rm E}_G$ \,this common elementary cycle 
language. For instance \,${\rm E}_{\rm Pair}\ =\ \{aa,b\}$. 
The elementary circularity implies the circularity.
\begin{lemma}\label{CircEleCirc}
Any elementary circular graph \,$G$ \,is circular with
\,${\rm L}_G\ \subseteq\ [\varepsilon]$\mbas{$E_G$}\,.
\end{lemma}
The converse of Lemma~\ref{CircEleCirc} is satisfied for any deterministic 
graph.
\begin{lemma}\label{CircCircEle}
Any deterministic circular graph \,$G$ \,is elementary circular with 
\,${\rm L}_G\ =\ [\varepsilon]$\mbas{$E_G$}\,.
\end{lemma}
Let us adapt Theorem~\ref{Main} by applying
Lemmas~\ref{CircSym}, \ref{CircEleCirc}, \ref{CircCircEle}.
\begin{theorem}\label{MainBis}
The Cayley graphs are the deterministic, strongly connected, simple graphs
which are equivalently vertex-transitive or circular or elementary circular.
\end{theorem}
Note that the strong connectedness condition in Theorem~\ref{MainBis} can not 
be simplified to the rootedness condition since the semi-line 
\,$\{\ n\ \fleche{a}\ n+1\ |\ n \geq 0\ \}$ \,is rooted, deterministic, simple, 
circular and elementary circular. 
However, this strongly connectedness may be replaced by the connectedness for
finite graphs. This is based on the property below.
\begin{lemma}\label{SymBasicFini}
Any finite connected vertex-transitive graph is strongly connected.
\end{lemma}
Let us adapt Theorem~\ref{MainBis} by applying Lemma~\ref{SymBasicFini}.
\begin{corollary}\label{MainTer}
The Cayley graphs of finite groups are the deterministic, connected, 
vertex-transitive, simple and finite graphs.
\end{corollary}

\subsection{Other Cayley graphs}

It is also sometimes defined the Cayley graph 
\,${\cal C}\inter{\mathsf{G},\mathsf{H}}$ \,for \,$\mathsf{H}$ \,a 
{\it weak generating subset} of \,$\mathsf{G}$ \,meaning that 
\,$\mathsf{H} \,\cup \,\mathsf{H}^{-1}$ \,is a generating subset of 
\,$\mathsf{G}$ \,where 
\,$\mathsf{H}^{-1} \,= \,\{\ h^{-1}\ |\ h \in \mathsf{H}\ \}$. 
In that case, we say that \,${\cal C}\inter{\mathsf{G},\mathsf{H}}$ \,is a 
{\it weak generated Cayley graph}. 
Theorem~\ref{MainBis} allows to characterize this more general family of graphs.
\begin{proposition}\label{Weak}
The weak generated Cayley graphs are the deterministic, co-deterministic, 
connected, simple and vertex-transitive graphs.
\end{proposition}
\proof
$\Longrightarrow$\,: \,Let \,$\mathsf{G}$ \,be a group and \,$\mathsf{H}$ 
\,be a non-empty weak generating subset of \,$\mathsf{G}$.\\
So \,${\cal C}\inter{\mathsf{G},\mathsf{H} \,\cup \,\mathsf{H}^{-1}}$ \,is a 
deterministic, co-deterministic, strongly-connected, vertex-transitive simple 
graph. 
By removing the arrows labeled in \,$\mathsf{H}^{-1}$, the graph 
\,${\cal C}\inter{\mathsf{G},\mathsf{H}}$ \,is connected and it remains 
deterministic, co-deterministic, simple and vertex-transitive.\\[0.25em]
$\Longleftarrow$\,: \,Let \,$G$ \,be a deterministic, co-deterministic, 
connected, simple, vertex-transitive graph. 
Let \,Inv \,be the set of labels \,$a$ \,such that \ 
$s\ \fleche{a}_G\ t \ \ \Longrightarrow \ \ t\ \fleche{}_G\ s$ \ for (all) 
\,$s,t \in V_G$ \,{\it i.e.}\\[0.25em]
\hspace*{8em}
${\rm Inv}\ =\ \{\ a \in A_G\ |\ \exists\ b \in A_G \ (ab \in {\rm L}_G)\ \}$.
\\[0.25em]
To each \,$a \in A_G - {\rm Inv}$, we associate a new element 
\,$\overline{a} \in A - A_G$ \,and we define the completion of \,$G$ \,by the
graph:\\[0.25em]
\hspace*{8em}$H\ =\ G \,\cup \,\{\ t\ \fleche{\overline{a}}\ s\ |\ 
s\ \fleche{a}_G\ t \,\wedge \,a \not\in {\rm Inv}\ \}$.\\[0.25em]
So \,$H$ \,is strongly connected and it remains simple,
deterministic and vertex-transitive.\\
By Theorem~\ref{MainBis}, \,$H$ \,is a Cayley graph:\\[0.25em]
\hspace*{15em}
$H\ =\ {\cal C}\inter{V_G,\mathsf{H}}$\\[0.25em]
for some generating subset \,$\mathsf{H}$ \,of a group \,$(V_G,\cdot)$ \,on
its vertex set.\\
Therefore \,$G$ \,is the weak generated Cayley graph:\\[0.25em]
\hspace*{8em}
$G\ =\ {\cal C}\inter{V_G,\mathsf{K}}$ \ \ with \ \ 
$\mathsf{K}\ =\ \{\ h \in \mathsf{H}\ |\ \inter{h} \in A_G\ \}$.
\qed\\[1em]
We also consider the Cayley graph \,${\cal C}\inter{\mathsf{G},\mathsf{H}}$ 
\,for every non-empty subset \,$\mathsf{H}$ \,of a group \,$\mathsf{G}$. 
Such graph is called here a {\it generalized Cayley graph}. 
These graphs form a more general class of graphs for which and under the 
assumption of the axiom of choice, we can give a simple graph characterization.
\begin{theorem}\label{MainFour}
Under ZFC set theory, the generalized Cayley graphs are the deterministic, 
co-deterministic, vertex-transitive simple graphs.
\end{theorem}
\proof
$\Longrightarrow$\,: \,By definition and for any subset \,$\mathsf{H}$ \,of 
\,$\mathsf{G}$, the graph \,${\cal C}\inter{\mathsf{G},\mathsf{H}}$ \,is 
simple, deterministic and co-deterministic. 
As already indicated  in the proof of Proposition~\ref{CaractCayley3}, this 
graph is vertex-transitive.\\[0.25em]
$\Longleftarrow$\,: \,Let \,$G$ \,be a deterministic, co-deterministic, 
vertex-transitive simple graph.\\
Let \,$\mathsf{Comp}$ \,be the set of connected components of \,$G$.\\
Using ZFC set theory, there exists a binary operation \,$\cdot$ \,such that
\,$(\mathsf{Comp},\cdot)$ \,is a group (in fact in ZF set theory, the axiom of
choice is equivalent to the property that any non-empty set has a group
structure).
We denoted by \,$I$ \,the identity element of \,$(\mathsf{Comp},\cdot)$.\\
Note that \,$I$ \,is connected, deterministic, co-deterministic, simple and
vertex-transitive.\\
By Proposition~\ref{Weak}, $I \,= \,{\cal C}\inter{V_I,\mathsf{H}}$ \,for some 
weak generating subset \,$\mathsf{H}$ \,of a group \,$(V_I,{\cdot}_I)$.\\
As \,$G$ \,is vertex-transitive, there is an isomorphism from\,$I$ \,to each
\,$C \in \mathsf{Comp}$. 
By the axiom of choice, we take for each \,$C \in \mathsf{Comp}$ \,an 
isomorphism \,$f_C$ \,from \,$I$ \,to \,$C$\,:\ $I \DoubleBigFleche{}_{f_C} C$.\\
It is assumed that \,$f_I$ \,is the identity on \,$V_I$\,.\\
We consider the group product \,$V_I{\croix}\mathsf{Comp}$, its subset 
\,$\mathsf{K} \,= \,\{\ (h,I)\ |\ h \in \mathsf{H}\ \}$ \,and the mapping 
\,$\inter{\ }$ \,defined by \,$\inter{(h,I)} \,= \,\inter{h}$ \,for any 
\,$h \in \mathsf{H}$\,.\\
We consider the bijection \,$f$ : \,$V_I{\croix}\mathsf{Comp}\ \fleche{}\ V_G$ 
\,defined by\\[0.25em]
\hspace*{10em}
$f(s,C) \,= \,f_C(s)$ \ for any \,$s \in V_I$ \,and \,$C \in \mathsf{Comp}$.
\\[0.25em]
We check (see Appendix) that 
\,${\cal C}\inter{V_I{\croix}\mathsf{Comp},\mathsf{K}} \,\DoubleBigArrow{}_f 
\,G$.
\qed

\section{Conclusion}

In this article, we have given simple graph conditions to characterize the
Cayley graphs as vertex-transitive graphs. 
This paper is only a first step in the study of symmetry to directed and
labelled graphs.

\newpage\noindent
\bibliographystyle{alpha}

\newpage\noindent

{\noindent}We give here proofs and complementary results.

\section{Standard results on automata}

By adding two new symbols \,$\iota,o$, every automaton 
\,${\cal A} \,= \,(G,I,F)$ \,can be seen as the coloured graph 
\,$G \,\cup \,\{\,(\iota,i)\,|\,i \in I\,\} \,\cup
\,\{\,(o,f)\,|\,f \in F\,\}$\,; \,we write \,$c\,s$ \,for any couple 
\,$(c,s)$ \,with \,$c \in \{\iota,o\}$ \,and \,$s \in V_G$\,.
We also write \,$V_{\cal A}$ \,for \,$V_G$ \,and \,$\fleche{}_{\cal A}$ \,for 
\,$\fleche{}_G$\,.\\
Let us start with a basic property on the path relation of a deterministic 
graph.\\[-2em]
\enonce{{\bf Lemma}~\ref{PathRes}.}
{For any \,$s\ \fleche{u}_G\ t$ \,and \,$F \subseteq V_G$ \,with \,$G$ 
\,deterministic, \,${\rm L}_G(t,F) \,= \,u^{-1}{\rm L}_G(s,F)$.}
By union, it is sufficient to check the equality for \,$F$ \,restricted to
a single vertex \,$f$.\\[0.25em]
$\subseteq$\,: \,Let \,$v \in {\rm L}_G(t,f)$ \,{\it i.e.}
\,$t\ \fleche{v}_G\ f$. 
So \,$s\ \fleche{uv}_G\ f$ \,hence \,$uv \in {\rm L}_G(s,f)$ \,{\it i.e.} 
\,$v \in u^{-1}{\rm L}_G(s,f)$.\\[0.25em]
$\supseteq$\,: \,Let \,$v \in u^{-1}{\rm L}_G(s,f)$ \,{\it i.e.}
\,$s\ \fleche{uv}_G\ f$.
There is \,$t'$ \,such that \,$s\ \fleche{u}_G\ t'\ \fleche{v}_G\ f$.\\
As \,$G$ \,is deterministic, $t = t'$ \,hence \,$v \in {\rm L}_G(t,f)$.
\qed\\[1em]
The simulation on automata involves the inclusion on the recognized languages.
\begin{lemma}\label{Simul}
If \ ${\cal A}$ \,is simulated by \,${\cal B}$ \ then \
${\rm L}({\cal A}) \,\subseteq \,{\rm L}({\cal B})$.
\end{lemma}
\proof
Let \,$R$ \,be a simulation from \,${\cal A}$ \,into \,${\cal B}$.
Let \,$u \in {\rm L}({\cal A})$.\\
There are \,$s,t \in V_{\cal A}$ \,such that \,$s\ \fleche{u}_{\cal A}\ t$ \,and 
\,$\iota\,s\,,\,o\,t \,\in \,{\cal A}$.\\
By definition of a simulation, there exists \,$p \in V_{\cal B}$ \,such that 
\,$\iota\,p \in {\cal B}$ \,and \,$s\ R\ p$.\\
By induction on \,$|u| \geq 0$, there exists \,$q \in V_{\cal B}$ \,such that
\,$p\ \fleche{u}_{\cal B}\ q$ \,and \,$t\ R\ q$.\\
As \,$o\,t \,\in \,{\cal A}$ \,and \,$t\ R\ q$, we have 
\,$o\,q \,\in \,{\cal B}$. Thus \,$u \in {\rm L}({\cal B})$.
\qed\\[1em]
Every automaton is reduced in all quotient.
\begin{lemma}\label{Reduc}
For every congruence \,$\equiva$ \,of any automaton \,${\cal A}$, \ 
${\cal A}\ \BigFleche{}_{\mbox{\tiny{\rm [}\ \,{\rm ]}}}\ 
[{\cal A}]\bas{\smallequiva}$\,.
\end{lemma}
\proof
The mapping \,$[\mbox{\ \ }]\bas{\smallequiva}\,:
\,s \in V_{\cal A}\ \mapsto\ [s] \in V_{[{\cal A}]\bas{\smallequiva}}$ \,is surjective.
\\
By definition of \,$[{\cal A}]$, if \,$s\ \fleche{a}_{\cal A}\ t$ \,then
\,$[s]\ \fleche{a}_{[{\cal A}]}\ [t]$.\\
Similarly, if \,$c\,s \in {\cal A}$ \,then \,$c\,[s] \in [{\cal A}]$.\\
Thus \,$[\mbox{\ \ }]\bas{\smallequiva}$ \,is a surjective morphism. 
It remains to check that \,$[\mbox{\ \ }]^{-1}\!\!\!\!\!\!\bas{\smallequiva}$ 
\,is a simulation.\\
If \,$[s]\ \fleche{a}_{[{\cal A}]}\ [t]$ \,then there exists 
\,$s'\ \fleche{a}_{\cal A}\ t'$ \,such that \,$s\ \equiva\ s'$ \,and 
\,$t\ \equiva\ t'$.\\
As \,$\equiva$ \,is a congruence, there is \,$t''$ \,such that 
\,$s\ \fleche{a}_{\cal A}\ t''$ \,and \,$t''\ \equiva\ t'$, hence 
\,$[t''] = [t]$.\\
If \,$\iota\,[s] \in [{\cal A}]$ \,then there is \,$s'$ \,such that 
\,$s\ \equiva\ s'$ \,and \,$\iota\,s' \in {\cal A}$.\\
If \,$o\,[s] \in [{\cal A}]$ \,then there is \,$s'$ \,such that
\,$s\ \equiva\ s'$ \,and \,$o\,s' \in {\cal A}$, hence \,$o\,s \in {\cal A}$.
\qed\\[1em]
The similarity between two automata corresponds to the reduction to a same
automaton, or to have the same minimal automaton.
\begin{lemma}\label{BisiMin}
For all automata \,${\cal A}$ \,and \,${\cal B}$, we have the following 
equivalences:\\[0.25em]
\hspace*{1.5em}a)\ \ ${\cal A}$ \,and \,${\cal B}$ \,are bisimilar,\\[0.25em]
\hspace*{1.5em}b)\ \ ${\cal A}$ \,and \,${\cal B}$ \,are reducible to a same 
automaton,\\[0.25em]
\hspace*{1.5em}c)\ \ ${\rm Min}({\cal A})$ \,and \,${\rm Min}({\cal B})$
\,are isomorphic.
\end{lemma}
\proof
${\bf a)\ \Longrightarrow\ c)}$\,: \,Let \,$R$ \,be a bisimulation between
\,${\cal A}$ \,and \,${\cal B}$. The relation
\begin{center}
$h \ = \ \{\ ([s]\bas{$\smallapprox_{\cal A}$}\,,[t]\bas{$\smallapprox_{\cal B}$})
\ |\ s\ R\ t\ \ \} \ = \
[\ \ ]^{-1}\!\!\!\!\!\!\Bas{$\smallapprox_{\cal A}$}\ \compose\ R\ 
\compose\ [\ \ ]\Bas{$\smallapprox_{\cal A}$}$
\end{center}
is a bisimulation between \,Min$({\cal A})$ \,and \,Min$({\cal B})$.\\
For that we have 
\,${\rm Min}({\cal A})\ \simeq_h\ {\rm Min}({\cal B})$, it 
remains to show that \,$h$ \,are \,$h^{-1}$ \,are injectives.\\
By symmetry, it is sufficient to show the injectivity of \,$h$.\\
Let 
\,$([s]\bas{$\smallapprox_{\cal A}$}\,,[t]\bas{$\smallapprox_{\cal B}$})\,,\,
([s]\bas{$\smallapprox_{\cal A}$}\,,[t']\bas{$\smallapprox_{\cal B}$}) \,\in \,h$. 
There exists \,$s_1\ R\ t_1$ \,and \,$s_2\ R\ t_2$ \,such that
\begin{center}
$[s_1]\bas{$\smallapprox_{\cal A}$} \,= \,[s]\bas{$\smallapprox_{\cal A}$} \,=
\,[s_2]\bas{$\smallapprox_{\cal A}$}$ \ \ \ ; \ \ \ 
$[t_1]\bas{$\smallapprox_{\cal B}$} \,= \,[t]\bas{$\smallapprox_{\cal B}$}$ \ \
\ ; \ \ \ 
$[t_2]\bas{$\smallapprox_{\cal B}$} \,= \,[t']\bas{$\smallapprox_{\cal B}$}$\,.
\end{center}
Therefore
\begin{center}
$(t_1,t_2) \ \in \ R^{-1}\ \compose\
\ [\ \ ]\Bas{$\smallapprox_{\cal A}$}\ \compose\ 
[\ \ ]^{-1}\!\!\!\!\!\!\Bas{$\smallapprox_{\cal A}$}\ \compose\ R$ \ \
is a bisimulation of \,${\cal B}$.
\end{center}
Thus \ $t_1\ \approx_{\cal B}\ t_2$ \ so \
$[t_1]\bas{$\smallapprox_{\cal B}$} \ = \,[t_2]\bas{$\smallapprox_{\cal B}$}$ \
{\it i.e.} \
$[t]\bas{$\smallapprox_{\cal B}$} \ = \,[t']\bas{$\smallapprox_{\cal B}$}$.
\\[0.25em]
${\bf c)\ \Longrightarrow\ b)}$\,: \,Let \ 
${\rm Min}({\cal A})\ \simeq_h\ {\rm Min}({\cal B})$.
From Lemma~\ref{Reduc}, we have
\begin{center}
${\cal A}\ \BigFleche{}_{[\ \ ]\bas{$h\ \compose\ \smallapprox_{\cal A}$}}\ 
{\rm Min}({\cal B})$ \ \ and \ \
${\cal B}\ \BigFleche{}_{[\ \ ]\bas{$\smallapprox_{\cal B}$}}\ {\rm Min}({\cal B})$.
\end{center}
${\bf b)\ \Longrightarrow\ a)}$\,:
\,Let \ ${\cal A}\ \BigFleche{}_g\ {\cal C}$ \
\ and \ \ ${\cal B}\ \BigFleche{}_h\ {\cal C}$.\\
So \,$g\ \compose\ h^{-1}$ \,is a bisimulation between \,${\cal A}$ \,and 
\,${\cal B}$.
\qed\\[1em]
The largest bisimulation for any deterministic and co-accessible automaton
coincides with the congruence of Nerode.
\enonce{{\bf Lemma}~\ref{Nerode}.}
{For any co-accessible automaton \,${\cal A} = (G,I,F)$ \,with \,$G$
\,deterministic,\\[0.25em]
\hspace*{8em}$s\ \approx_{\cal A}\ t \ \ \ \Longleftrightarrow \ \ \ 
{\rm L}_G(s,F) \,= \,{\rm L}_G(t,F)$ \ \ for all \,$s,t \in V_{\cal A}$\,.}
{\bf i)} Let \,$R$ \,be a bisimulation of any automaton 
\,${\cal A} = (G,I,F)$. Let \,$s\ R\ t$.\\
So the automata \,$(G,s,F)$ \,and \,$(G,t,F)$ \,are bisimilar. 
By Lemma~\ref{Simul}, we get
\begin{center}
$s\ R\ t \ \ \ \Longrightarrow \ \ \ {\rm L}_G(s,F) \,= \,{\rm L}_G(t,F)$
\end{center}
hence the direct implication of this lemma for \ $R$ \,equal to
\,$\approx_{\cal A}$\,.\\[0.25em]
{\bf ii)} Assume that \,${\cal A}$ \,is co-accessible with \,$G$
\,deterministic. We define the relation
\begin{center}
$R \ = \ \{\ (s,t)\ |\ s,t \in V_G \,\wedge 
\,{\rm L}_G(s,F) \,= \,{\rm L}_G(t,F)\ \}$.
\end{center}
For the property \,$R \,\subseteq\ \ \approx_{\cal A}$\,, it suffices to show
that \,$R$ \,is a bisimulation.\\
As \,$R$ \,is an equivalence hence symmetric, it is sufficient to check that 
\,$R$ \,is a simulation.\\
As \,$R$ \,is reflexive, its domain is \,$V_G$ \,and for
\,$\iota\,s \in {\cal A}$, we have \,$s\ R\ s$. Furthermore
\begin{center}
$o\,s \in {\cal A} \ \ \Longleftrightarrow \ \ \varepsilon \in {\rm L}_G(s,F)$
\ \ \ hence \ \ \ $(s\ R\ t \,\wedge \,o\,s \in{\cal A} ) \ \ \Longrightarrow
\ \ o\,t \in {\cal A}$.
\end{center}
Finally, let \,$s\ R\ t$ \ and \ $s\ \fleche{a}_G\ s'$.\\
As \,$G$ \,is deterministic and by Lemma~\ref{PathRes}, 
${\rm L}_G(s',F) \,= \,a^{-1}{\rm L}_G(s,F) \,= \,a^{-1}{\rm L}_G(t,F)$.\\
As \,${\cal A}$ \,is co-accessible, \,${\rm L}_G(s',F) \,\neq \,\emptyset$
\ hence  \ $a^{-1}{\rm L}_G(t,F) \,\neq \,\emptyset$.\\
So there is \,$t' \in V_G$ \,such that \,$t\ \fleche{a}_G\ t'$.\\
As \,$G$ \,is deterministic and by Lemma~\ref{PathRes}, 
${\rm L}_G(t',F) \,= \,a^{-1}{\rm L}_G(t,F) \,= \,{\rm L}_G(s',F)$ \ {\it i.e.} 
\ $s'\ R\ t'$.
\qed\\[1em]
Every co-deterministic and co-accessible automaton is minimal and remains
minimal by determinization.
\enonce{{\bf Lemma}~\ref{CoDetMin}.}
{For any automaton \,${\cal A}$ \,co-deterministic and co-accessible,\\
\hspace*{5.5em}${\cal A}$ \,and \,${\rm Det}({\cal A})$ \,are minimal.}
Let us denote \,${\cal A} = (G,I,\{f\})$ \,and 
\,${\rm Det}({\cal A}) = (H,\{I\},F)$.\\
{\bf i)} Let us check that \,${\cal A}$ \,is minimal. 
Let \,$R$ \,be a bisimulation of \,${\cal A}$\,. Let \,$s\,R\,t$.\\
So \,${\rm L}_G(s,f) \,= \,{\rm L}_G(t,f) \,\neq \,\emptyset$ \,since
\,${\cal A}$ \,is co-accessible. 
As \,${\cal A}$ \,is co-deterministic, $s = t$.\\[0.25em]
{\bf ii)} For any vertex \,$J$ \,of \,$H$ \,{\it i.e.} 
\,$J = I{\cdot}v \neq \emptyset$ \,for some \,$v \in A^*$, we have
\begin{center}
${\rm L}_H(J,F)\ =\ \{\ u \in A^*\ |\ J{\cdot}u \in F\ \}\ =\ 
\{\ u \in A^*\ |\ f \in J{\cdot}u\ \}\ =\ {\rm L}_G(J,f)$.
\end{center}
In particular \,${\rm L}({\rm Det}({\cal A}))\ =\ {\rm L}_H(I,F)\ =\
{\rm L}_G(I,F)\ =\ {\rm L}({\cal A})$.\\[0.25em]
{\bf iii)} Let us check that \,${\rm Det}({\cal A})$ \,is minimal.
Consider two bisimilar vertices \,$I{\cdot}u,I{\cdot}v$ \,of
\,${\rm Det}({\cal A})$.\\
In particular \,${\rm L}_H(I{\cdot}u,F) \,= \,{\rm L}_H(I{\cdot}v,F)$.
Thus by $(ii)$, \,${\rm L}_G(I{\cdot}u,f) \,= \,{\rm L}_G(I{\cdot}v,f)$.\\
Let \,$s \in I{\cdot}u$.
As \,${\cal A}$ \,is co-accessible, there is 
\,$w \in {\rm L}_G(s,f) \,\subseteq \,{\rm L}_G(I{\cdot}u,f) \,=
\,{\rm L}_G(I{\cdot}v,f)$.\\
There exists \,$t \in I{\cdot}v$ \,such that \,$w \in {\rm L}_G(t,f)$.\\
As \,${\cal A}$ \,is co-deterministic, $s = t$.
Thus \,$I{\cdot}u \subseteq I{\cdot}v$ \,and by symmetry
\,$I{\cdot}u = I{\cdot}v$.
\qed\\[1em]
By definition, ${\rm Can}(L)$ \,is deterministic and reduced; its recognized 
language is \,$L$.
\begin{lemma}\label{ResBasic}
For any language \,$L$, the automaton \,${\rm Can}(L)$ \,is deterministic, 
reduced\\
\hspace*{6.5em}and recognizes \,$L$.
\end{lemma}
\proof
We just have to check that \,$L \,= \,{\rm L}({\rm Can}(L))$.\\
$\subseteq$\,: \ Let \,$u \in L$.
Then we have \ $L\ \fleche{u}_{{\rm Can}(L)}\ u^{-1}L$ \ \ {\it i.e.} \ 
$u \in {\rm L}({\rm Can}(L))$.\\[0.25em]
$\supseteq$\,: \ Let \,$u \in {\rm L}({\rm Can}(L))$. 
There is \,$v \in L$ \,such that \,$u \in {\rm L}_{{\rm Can}(L)}(L,v^{-1}L)$.\\
It follows that \ $\varepsilon \in v^{-1}L$ \ and \ 
$L\ \fleche{u}_{{\rm Can}(L)}\ v^{-1}L$.\\
As \,${\rm Can}(L)$ \,is deterministic, we get \,$u^{-1}L = v^{-1}L$. 
So \,$\varepsilon \in u^{-1}L$ \ {\it i.e.} \,$u \in L$.
\qed\\[1em]
The automaton \,${\rm Can}(L)$ \,is the unique minimal, deterministic and
reduced automaton recognizing \,$L$.
\enonce{{\bf Lemma}~\ref{MinDet}.}
{For any deterministic and reduced automaton \,${\cal A}$, the automaton 
\,${\rm Min}({\cal A})$ \,is\\
\hspace*{5.5em}isomorphic to \,${\rm Can}({\rm L}({\cal A}))$.}
Let \,${\cal A} = (G,i,F)$. As \,${\cal A}$ \,is deterministic and
co-accessible, and by Lemma~\ref{Nerode},\\[0.25em]
\hspace*{6em}$s\ \approx_{\cal A}\ t \ \ \ \Longleftrightarrow \ \ \ 
{\rm L}_G(s,F) \,= \,{\rm L}_G(t,F)$ \ \ for all \,$s,t \in V_{\cal A}$\,.
\\[0.25em]
Let \,$L \,= \,{\rm L}({\cal A}) \,= \,{\rm L}_G(i,F)$. 
As \,${\cal A}$ \,is deterministic and by Lemma~\ref{PathRes},\\[0.25em]
\hspace*{6em}${\rm L}_G(s,F) \,= \,u^{-1}L$ \ \ for all \,$i\ \fleche{u}_G\ s$.
\\[0.25em]
As \,${\cal A}$ \,is accessible and co-accessible, we can define the mapping
from \,$V_{{\rm Min}({\cal A})}$ \,into \,$V_{{\rm Can}({\rm L}({\cal A}))}$~by\\[0.25em]
\hspace*{6em}$[s]\bas{$\smallapprox_{\cal A}$}\ \longmapsto\ u^{-1}L$ \ \ for \ 
$i\ \fleche{u}_G\ s$\\[0.25em]
which is an isomorphism from \,${\rm Min}({\cal A})$ \,to 
\,${\rm Can}({\rm L}({\cal A}))$.
\qed\\[1em]
Due to the importance of Corollary~\ref{EquivRes}, we give a direct proof.
\enonce{{\bf Corollary}~\ref{EquivRes}.}
{For any graph \,$G$ \,strongly connected, deterministic and co-deterministic,\\
\hspace*{6.5em}$G$ \,is isomorphic to \,$\overrightarrow{{\rm L}_G(s,t)}$
\,for all \,$s,t \in V_G$\,.}
Let \,$G$ \,be a strongly connected, deterministic and co-deterministic graph.\\
Let \,$s,t \in V_G$ \,and \,$L = {\rm L}_G(s,t)$. 
Let us check that for all \,$r \in V_G$ \,and \,$u \in A^*$,
\begin{equation}\label{Equiv}
\hspace*{8em}{\rm L}_G(r,t)\ =\ u^{-1}L \ \ \Longleftrightarrow \ \ 
s\ \fleche{u}_G\ r.
\end{equation}
\hspace*{1em}$\Longleftarrow$\,: \,as \,$G$ \,is deterministic, it suffices to
apply Lemma~\ref{PathRes}.\\[0.25em]
\hspace*{1em}$\Longrightarrow$\,: \,there exists \,$v$ \,such that \,$uv \in L$
\,and \,$r\ \fleche{v}\ t$.\\
\hspace*{3em}As \,$uv \in L$, there exists \,$r'$ \,such that 
\,$s\ \fleche{u}\ r'\ \fleche{v}\ t$.\\
\hspace*{3em}As \,$G$ \,is co-deterministic, we get \,$r = r'$ \,hence 
\,$s\ \fleche{u}\ r$.\\[0.25em]
Thus Property (\ref{Equiv}) is checked. Let \,$H = \overrightarrow{L}$.\\
We now show that \,$G$ \,is isomorphic to \,$H$ \,according to the following 
mapping\,:
\begin{center}
\begin{tabular}{lccl}
$h :$ & $V_G$ & $\longrightarrow$ & 
$\{\ u^{-1}L\ |\ u \in A^*\ \} \,- \,\{\emptyset\}$\\[0.25em]
 & $r$ & $\longmapsto$ & ${\rm L}_G(r,t)$
\end{tabular}
\end{center}
Let us check that \,$h$ \,is well-defined. Let \,$r \in V_G$\,.\\
As \,$G$ \,is strongly connected, ${\rm L}_G(r,t) \neq \emptyset$ \,and there 
exists \,$u \in A^*$ \,such that \,$s\ \fleche{u}_G\ r$.\\
By~(\ref{Equiv}), we have \,${\rm L}_G(r,t)\ =\ u^{-1}L$.\\[0.25em]
Let us check that \,$h$ \,is surjective. 
Let \,$u \in A^*$ \,such that \,$u^{-1}L \neq \emptyset$.\\
Thus there exists \,$r$ \,such that \,$s\ \fleche{u}_G\ r$. 
By~(\ref{Equiv}), we have \,${\rm L}_G(r,t)\ =\ u^{-1}L$.\\[0.25em]
Let us check that \,$h$ \,is injective.\\
Let \,$p \neq q \in V_G$\,. As \,$G$ \,is co-deterministic, we have
\begin{center}
$h(p) \,\cap \,h(q) \ = \ {\rm L}_G(p,t) \,\cap \,{\rm L}_G(q,t) \ = \ 
\emptyset$.
\end{center}
As \,$h(p),h(q) \neq \emptyset$, we get \,$h(p) \neq h(q)$. 
So \,$h$ \,is a bijection.\\
To show that \,$h$ \,is an isomorphism from \,$G$ \,to \,$H$, it remains to 
verify that
\begin{center}
$r\ \fleche{a}_G\ r' \ \ \ \Longleftrightarrow \ \ \ h(r)\ \fleche{a}_H\ h(r')$.
\end{center}
If \ $r\ \fleche{a}_G\ r'$ \ then by Lemma~\ref{PathRes},\ 
${\rm L}_G(r',t) = a^{-1}{\rm L}_G(r,t)$ \ hence
\begin{center}
$h(r) \,= \,{\rm L}_G(r,t)\,\ \fleche{a}_H\,\ {\rm L}_G(r',t) \,= 
\,h(r')$.
\end{center}
If \ $h(r)\ \fleche{a}_H\ h(r')$ \ then \
${\rm L}_G(r',t) \,= \,a^{-1}{\rm L}_G(r,t)$.\\
As \,$G$ \,is strongly connected, there exists \,$u \in A^*$ \,such that 
\,$s\ \fleche{u}\ r$. By~(\ref{Equiv})
\begin{center}
${\rm L}_G(r',t) \ = \ a^{-1}{\rm L}_G(r,t) \ = \ a^{-1}(u^{-1}L) \ = \ 
(ua)^{-1}L$.
\end{center}
By~(\ref{Equiv}), \,$s\ \fleche{ua}\ r'$ \,and as \,$G$ \,is deterministic, 
we get \,$r\ \fleche{a}\ r'$.
\qed\\[1em]
Co-determinism in the statement of Corollary~\ref{EquivRes} is required. 
Indeed, the following graph~\,$G$\,:
\begin{center}
\includegraphics{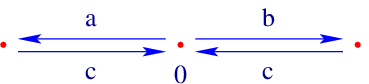}
\end{center}
is deterministic, strongly connected and one of its path language is 
\,${\rm L}_G(0,0) \,= \,((a+b)c)^*$ \,whose following canonical graph
\begin{center}
\includegraphics{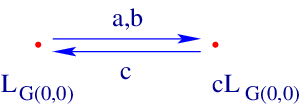}
\end{center}
is not isomorphic to \,$G$.

\section{Cayley graphs}

By removing labels and orientations of arrows of a graph \,$G$, we obtain its 
{\it skeleton}
\begin{center}
${\rm Ske}(G) \ = \ \{\ \{s,t\}\ |\ s\ \fleche{}_G\ t\ \}$
\end{center}
which is a non-directed and unlabelled graph, or equivalently a bi-directed
graph labelled by a single symbol \,$\diese$\,:
\,$\{\ s\ \fleche{\smalldiese}\ t\ |\ \exists\ a\ (s\ \fleche{a}_G\ t \,\vee 
\,t\ \fleche{a}_G\ s)\ \}$.\\
The skeleton of \,Even \,is represented by
\begin{center}
\includegraphics{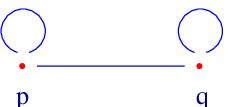}
\end{center}
\noindent Sabidussi's theorem gives a characterization of skeletons of the 
Cayley graphs.
\begin{theorem}\label{Sabidussi}
{\rm (Sabidussi)} \ For any non-directed and unlabelled graph \,$G$,\\[0.25em]
\hspace*{1em}$G$ \,is the skeleton of a Cayley graph \ if and only if\\[0.25em]
\hspace*{1em}$G$ \,is connected and there is a subgroup of \,${\rm Aut}(G)$ 
\,with a free transitive action on \,$G$.
\end{theorem}
We can removed the connectedness condition of this theorem by extending the 
definition of a Cayley graph for any non-empty subset \,$\mathsf{H}$ \,of a
group \,$\mathsf{G}$. However, this condition is necessary when
\,$\mathsf{H}$ \,is a generating subset of \,$\mathsf{G}$. 
For instance, consider the non-connected graph 
\,$G \,= \,\{1\ \mbox{---}\ 2\,,\,3\ \mbox{---}\ 4\}$.
A subgroup of \,${\rm Aut}(G)$ \,is the Klein group of permutation 
representation:\\[-2em]\mbox{}
\begin{center}
$\mathsf{K}_4 \,= \,\{()\,,\,(1,2)(3,4)\,,\,(1,3)(2,4)\,,\,(1,4)(2,3)\}$
\end{center}
whose mapping \,$(f,s) \,\mapsto \,f(s)$ \,is a free and transitive action on 
\,$G$.\\
Note that the connectedness of a graph \,$G$ \,is a simple graph property and 
it is quite more difficult to know whether there exists a subgroup of
\,${\rm Aut}(G)$ \,with a free and transitive action on \,$G$. 
For instance, the following {\it Petersen skeleton}:
\begin{center}
\includegraphics{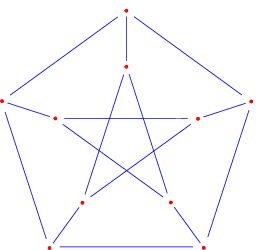}
\end{center}
is connected, vertex-transitive but it has no free transitive group action.
\\[1em]
Let us conclude the proof of Proposition~\ref{CaractCayley3}
\enonce{{\bf Proposition}~\ref{CaractCayley3}.}
{A graph \,$G$ \,is isomorphic to a Cayley graph of a group \,$\mathsf{G}$ \,if 
and only if\\
$G$ \,is a deterministic rooted simple graph with a free transitive action of 
\,$\mathsf{G}$ \,on \,$G$.}
Let us end this proof by checking that 
\,$\overline{G} \,= \,{\cal C}\inter{\mathsf{G},\mathsf{H}}$ \,and 
\,$\mathsf{H}$ \,is a generating subset of \,$\mathsf{G}$.\\[0.25em]
Let \,$p\ \fleche{a}_G\ q$. As $\bouboule$ \,is transitive, there is 
\,$g \in \mathsf{G}$ \,such that \,$g{\bouboule}p \,= \,r$.\\
As $\bouboule$ \,is a morphism, we get \,$r \,= \,g{\bouboule}p\ \fleche{a}_G\ 
g{\bouboule}q$. We denote \,$s \,= \,g{\bouboule}q$.\\
So \,$p \,= \,g^{-1}{\bouboule}r$ \,{\it i.e.} \,$g^{-1} = g_p$\,.\\
Moreover \,$a = \inter{g_s}$ \,and 
\,$q \,= \,g^{-1}{\bouboule}s \,= \,g_p{\bouboule}s \,= \,(g_pg_s){\bouboule}r$ 
\,{\it i.e.} \,$g_q \,= \,g_pg_s$\,.\\ 
Thus \,$A_G = \inter{\mathsf{H}}$ \,and was obtained the necessary condition
of the following equivalence
\begin{equation}\label{Equiv0}
\hspace*{10em}p\ \fleche{\interFootnote{g_s}}_G\ q \ \ \ \Longleftrightarrow \ 
\ \ (g_q \,= \,g_pg_s \,\wedge \,g_s \in \mathsf{H})\,.
\end{equation}
For the sufficient condition of (\ref{Equiv0}), let \,$g_q \,= \,g_pg_s$ 
\,with \,$g_s \in \mathsf{H}$.\\
So \,$q \,= \,g_q{\bouboule}r \,= \,g_pg_s{\bouboule}r \,= \,g_p{\bouboule}s$. 
As \,$r\ \fleche{\interFootnote{g_s}}_G\ s$, we get 
\,$p \,= \,g_p{\bouboule}r\ \fleche{\interFootnote{g_s}}_G\ g_p{\bouboule}s \,= \,q$.
\\[0.25em]
Therefore and by (\ref{Equiv0}), we get 
\,$\overline{G} \,= \,{\cal C}\inter{\mathsf{G},\mathsf{H}}$ \,because
\\[0.25em]
\hspace*{10em}
$p\ \fleche{\interFootnote{g_s}}_G\ q \ \ \ \Longleftrightarrow \ \ \ (g_q 
\,= \,g_pg_s \,\wedge \,g_s \in \mathsf{H})) \ \ \ \Longleftrightarrow \ \ \ 
g_p\ \fleche{\interFootnote{g_s}}_{{\cal C}\interFootnote{\mathsf{G},\mathsf{H}}}\ 
g_q$\,.\\[0.25em]
It remains to check that \,$\mathsf{H}$ \,is a generating subset of 
\,$\mathsf{G}$.\\
Let \,$g \in \mathsf{G}$. As \,$r$ \,is a root of \,$G$, there is a path from 
\,$r$ \,to \,$g{\bouboule}r$\,:\\[0.25em]
\hspace*{10em}
$r \,= \,r_0\ \fleche{a_1}_G\ r_1 \ldots r_{n-1}\ \fleche{a_n}_G\ r_n \,= 
\,g{\bouboule}r$.\\[0.25em]
As \,$A_G = \inter{\mathsf{H}}$, there are (unique) 
\,$g_{s_1},\ldots,g_{s_n} \in \mathsf{H}$ \,such that 
\,$a_1 = \inter{g_{s_1}},\ldots,a_n = \inter{g_{s_n}}$\,.\\
By (\ref{Equiv0}), we have 
\,$g_{r_1} \,= \,g_{r_0}g_{s_1},\ldots,g_{r_n} \,= \,g_{r_{n-1}}g_{s_n}$ \,hence 
\,$g \,= \,g_{r_n} \,= \,g_{s_1}{\ldots}g_{s_n}$\,.
\qed\\[1em]
The Cayley graph of a language \,$L$ \,has for cycle language the closure of
\,$L$ \,by adding and deleting factors of \,$L$.
\enonce{{\bf Lemma}~\ref{LangCayley}.}
{For any group presentation language \,$L$, 
\,${\rm L}_{{\cal C}(L)} \,= \,[\varepsilon]$\mbas{$L$} \,and
\,${\cal C}(L) \,= \,{\cal C}([\varepsilon]$\mbas{$L$}$)$.}
We have \ ${\rm L}_{{\cal C}(L)} \,= 
\,{\rm L}({\cal C}(L),[\varepsilon]$\mbas{$L$}$,[\varepsilon]$\mbas{$L$}) \ 
and for any \,$u \in A_L^*$\,,
\begin{center}
$[\varepsilon]$\mbas{$L$}\ $\fleche{u}_{{\cal C}(L)}\ [\varepsilon]$\mbas{$L$} \ 
\ \ $\Longleftrightarrow \ \ \ [u]$\mbas{$L$}$ \,= \,[\varepsilon]$\mbas{$L$} 
\ \ \ $\Longleftrightarrow \ \ \ u \in [\varepsilon]$\mbas{$L$}\,.
\end{center}
As \,$\fleche{}_{[\varepsilon]\mbas{$L$}} \,\subseteq \,\gene{*}{}$$_L$ \,with 
\,$L \subseteq [\varepsilon]$\mbas{$L$}\,, we get
\,$\gene{*}{}$$_{[\varepsilon]\mbas{$L$}} \,= \,\gene{*}{}$$_L$\,.\\
Thus \,$[\varepsilon]$\mbas{$L$} \,is a group presentation language and
\,${\cal C}(L) \,= \,{\cal C}([\varepsilon]$\mbas{$L$}$)$.
\qed\\[1em]
The stability for a group presentation language \,$L$ \,means that its Cayley
graph \,${\cal C}(L)$ \,has for cycle language \,$L$.
\enonce{{\bf Lemma}~\ref{LangStable}.}
{A non-empty language \,$L$ \,is stable \,if and only if 
\,$L \,= \,[\varepsilon]$\mbas{$L$}\,.}
For every \,$u \in L$, 
\,$u\ \fleche{}$$_L\ \varepsilon$ \,hence 
\,$u \in [u]$\mbas{$L$} \,$= \,[\varepsilon]$\mbas{$L$}\,. 
Thus \,$L \subseteq [\varepsilon]$\mbas{$L$}.\\
$\Longrightarrow$\,: \,Assume that \,$L$ \,is a non-empty stable language. 
It remains to check that \,$[\varepsilon]$\mbas{$L$} $\subseteq L$.\\
There exists \,$u \in L$. We get by stability \,$\varepsilon \in L$.\\
By induction on \,$n \geq 0$, 
\,$(\varepsilon \,\gene{}{}$$_L$ $u_1 \ldots 
\,\gene{}{}$$_L u_n)\ \Longrightarrow\ u_n \in L$. 
Thus \,$[\varepsilon]$\mbas{$L$} $\subseteq L$.\\[0.25em]
$\Longleftarrow$\,: \,Assume that \,$L \,= \,[\varepsilon]$\mbas{$L$}\,.\\ 
Let \,$u \,\gene{}{}$$_L \,v$ \,with \,$u \in L$. 
It remains to verify that \,$v \in L$.\\
By hypothesis \,$u \in [\varepsilon]$\mbas{$L$} \,hence 
\,$v \in [u]$\mbas{$L$} \,$= \,[\varepsilon]$\mbas{$L$} \,$= \,L$.
\qed

\section{Elementary graph properties}

We establish basic properties for vertex-transitive and circular graphs.\\
Note that the following graph:
\begin{center}
\includegraphics{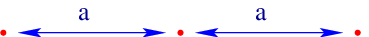}
\end{center}
is strongly connected, circular but is not vertex-transitive.\\
A first property for vertex-transitive graphs is that the strong connectedness 
is reduced to the existence of a root.
\enonce{{\bf Lemma}~\ref{RootTransitive}.}
{Any rooted vertex-transitive graph is strongly connected.}
Let \,$G$ \,be a vertex-transitive graph with root \,$r$.\\
Let us check that \,$G$ \,is strongly connected. Let \,$s \in V_G$\,.\\
It is sufficient to check that \,$s\ \fleche{}_G^*\ r$.\\
As \,$G$ \,is vertex-transitive, there exists an automorphism \,$h$ \,of \,$G$ 
\,such that \,$h(r) = s$.\\
As \,$r$ \,is a root, we have \,$r\ \fleche{}^*\ h^{-1}(r)$ \ hence \
\,$s = h(r)\ \fleche{}^*\ r$.
\qed\\[1em]
In addition, any deterministic circular graph with a root is co-deterministic.
\enonce{{\bf Lemma}~\ref{CoDet}.}
{Any deterministic and strongly connected circular graph is co-deterministic.}
Let \,$G$ \,be a deterministic and strongly connected circular graph.\\
Let \,$x\ \fleche{a}\ z$ \,and \,$y\ \fleche{a}\ z$.\\
As \,$G$ \,is strongly connected, there exists \,$u \in A^*$ \,such that  
\,$z\ \fleche{u}\ x$.\\
As \,$G$ \,is circular, we have \ 
$au \,\in \,{\rm L}_G(x,x) \,= \,{\rm L}_G(y,y)$.\\
As \,$G$ \,is deterministic, we get \ $z\ \fleche{u}\ y$. 
Thus \,$z\ \fleche{u}\ x$ \,and \,$z\ \fleche{u}\ y$.\\
As \,$G$ \,is deterministic, \,$x = y$.
\qed\\[1em]
The cycle language of any circular graph is stable when the graph is
deterministic.
\enonce{{\bf Lemma}~\ref{ClasseVide}.}
{For any deterministic circular graph \,$G$, \,${\rm L}_G$ \,is a stable 
language.}
Let us show that \,${\rm L}_G$ \,is preserved by self-insertion.\\
Let \,$uw,v \in {\rm L}_G$\,. One checks that \,$uvw \in {\rm L}_G$\,.\\
There exists \,$s,t \in V_G$ \,such that 
\,$s\ \fleche{u}_G\ t\ \fleche{w}_G\ s$.\\
As \,$v \in {\rm L}_G$\,, we have \,$t\ \fleche{v}\ t$ \ therefore \ 
$uvw \in {\rm L}_G(s,s) = {\rm L}_G$\,.\\
Let us show that \,${\rm L}_G$ \,is preserved by self-deletion.\\
Let \,$uvw,v \in {\rm L}_G$\,. One checks that \,$uw \in {\rm L}_G$\,.\\
There exists \,$r,s,t \in V_G$ \,such that 
\,$r\ \fleche{u}_G\ s\ \fleche{v}_G\ t\ \fleche{w}_G\ r$.\\
As \,$v \in {\rm L}_G$ \,we have \,$s\ \fleche{v}_G\ s$. 
As \,$G$ \,is deterministic, we get \,$s = t$.\\
So \,$uw \in {\rm L}_G(r,r) = {\rm L}_G$\,.
\qed\\[1em]
The cycle language of any circular graph is closed under conjugacy when the
graph is strongly connected.
\enonce{{\bf Lemma}~\ref{SymFort}.}
{For any strongly connected circular graph \,$G$,\\[0.25em]
\hspace*{6em}${\rm L}_G$ \,is closed under conjugacy and of letter set 
\,$A_{{\rm L}_G} \,= \,A_G$\,.}
Let us check that \,${\rm L}_G$ \,is closed under conjugacy.
Let \,$uv \in {\rm L}_G$\,.\\
There exists \,$s,t$ \,such that \,$s\ \fleche{u}\ t\ \fleche{v}\ s$. 
So \,$vu \in {\rm L}_G(t,t) = {\rm L}_G$\,.\\
We have \,$A_{{\rm L}_G} \subseteq A_G$\,. Let us check the reverse inclusion.
Let \,$a \in A_G$\,.\\
There exists \,$s,t$ \,such that \,$s\ \fleche{a}\ t$.\\
As \,$G$ \,is strongly connected, there exists \,$u$ \,such that 
\,$t\ \fleche{u}\ s$.\\
Thus \,$au \in {\rm L}_G$ \,hence \,$a \in A_{{\rm L}_G}$\,.
\qed\\[1em]
The cycle language of any circular graph is a group presentation when the graph 
is deterministic, simple with a root.
\enonce{{\bf Lemma}~\ref{CircPres}.}
{For any strongly connected, deterministic circular simple graph \,$G$,
\\[0.25em]
\hspace*{6em}${\rm L}_G$ \,is a group presentation language.}
By Lemma~\ref{ClasseVide}, \,${\rm L}_G$ \,is stable hence by
Lemma~\ref{LangStable}, \,$[\varepsilon]$\mbas{${\rm L}_G$} \,$=
\,{\rm L}_G$\,.\\
By Lemma~\ref{SymFort}, \,${\rm L}_G$ \,is closed under conjugacy and of
letter set \,$A_{{\rm L}_G} \,= \,A_G$\,.\\
As \,$A_G \neq \emptyset$, \,${\rm L}_G$ \,satisfies condition~$(i)$ \,of a 
group presentation language.\\
Let \,$a \in A_{{\rm L}_G}$\,. \,There exists words \,$u,v$ \,such that
\,$uav \in {\rm L}_G$\,.
Thus \,$avu, vua \in {\rm L}_G \,= \,[\varepsilon]$\mbas{${\rm L}_G$}\,.\\
So \,${\rm L}_G$ \,satisfies condition~$(ii)$ \,of a group presentation
language.\\
Let us check condition $(iii)$.
Let \,$a,b \in A_{{\rm L}_G}$ \,such that
\,$[a]$\mbas{${\rm L}_G$} \,$= \,[b]$\mbas{${\rm L}_G$}\,.
So \,$a\ \gene{*}{}$$_{{\rm L}_G}\ b$.\\
By \,$(ii)$, there exists \,$u \in A^*$ \,such that
\,$ua \in [\varepsilon]$\mbas{${\rm L}_G$}\,.
So \,$ub \in [\varepsilon]$\mbas{${\rm L}_G$}\,.
Thus \,$ua,ub \in {\rm L}_G$\,.\\
Let \,$r \in V_G$\,. There are \,$s,t$ \,such that
\,$r\ \fleche{u}_G\ s\ \fleche{a}_G\ r$ \,and
\,$r\ \fleche{u}_G\ t\ \fleche{b}_G\ r$.\\
As \,$G$ \,is deterministic, \,$s = t$. As \,$G$ \,is simple, $a = b$.\\
Thus \,${\rm L}_G$ \,is a group presentation language.
\qed\\[1em]
The concepts of circularity and vertex-transitivity coincide for deterministic
and strongly connected graphs.
\enonce{{\bf Lemma}~\ref{CircSym}.}
{For any deterministic and strongly connected graph \,$G$,\\
\hspace*{8em}$G$ \,is vertex-transitive if and only if \ $G$ \,is circular.}
Let \,$G$ \,be a strongly connected, deterministic and circular graph.\\
We have to check that \,$G$ \,is vertex-transitive.\\
This follows from the fact that the equivalence of vertices corresponds to the 
equality of their cycle language:
\begin{center}
$s\ \simeq_G\ t \ \ \ \Longleftrightarrow \ \ \ 
{\rm L}_G(s,s) \,= \,{\rm L}_G(t,t)$.
\end{center}
The necessary condition is immediate. Let us show the sufficient condition.\\
Let \,$s,t \in V_G$ \,such that \,${\rm L}_G(s,s) \,= \,{\rm L}_G(t,t)$.\\
We define the relation
\begin{center}
$k \ = \ \{\ (s',t')\ |\ \exists\ u \in A^* \ 
(s\ \fleche{u}_G\ s' \,\wedge \,t\ \fleche{u}_G\ t')\ \}$
\end{center}
We have \,$(s,t) \in k$. Let us prove that \,$k$ \,is an automorphism of \,$G$.
\\
Let us check that The domain of \,$k$ \,is \,$V_G$\,. 
Let \,$s' \in V_G$\,.\\
As \,$G$ \,is strongly connected, we have \,$s\ \fleche{u}\ s'$ \,for some
\,$u \in A_G^*$\,.\\
As \,$G$ \,is deterministic, \,${\rm L}_G(s',s) \,= \,u^{-1}{\rm L}_G(s,s) \,=
\,u^{-1}{\rm L}_G(t,t)$.\\
As \,$G$ \,is strongly connected, \,${\rm L}_G(s',s) \neq \emptyset$
\,{\it i.e.} \,$u^{-1}{\rm L}_G(t,t) \neq \emptyset$.\\
So there exists \,$t'$ \,such that \,$t\ \fleche{u}\ t'$.\\
In addition, \,$k$ \,is a mapping since for any \,$(p,q)\,,\,(p,r) \in k$,
there exists \,$u,v \in A^*$ \,such that
\begin{center}
$s\ \fleche{u}\ p\ \ ,\ \ t\ \fleche{u}\ q\ \ ,\ \ 
s\ \fleche{v}\ p\ \ ,\ \ t\ \fleche{v}\ r$.
\end{center}
As \,$G$ \,is strongly connected, there exists \,$w \in A^*$ \,such that 
\,$p\ \fleche{w}\ s$.\\
Also \,$uw,vw \,\in {\rm L}_G(s,s) \,= \,{\rm L}_G(t,t)$.\\
As \,$G$ \,is deterministic, we have 
\,$q\ \fleche{w}\ t$ \,and \,$r\ \fleche{w}\ t$.\\
As \,$G$ \,is co-deterministic, we get \,$q = r$.\\
By symmetry of \,$s,t$, the relation \,$k^{-1}$ \,is a mapping, hence 
\,$k$ \,is a bijection.\\
By symmetry of \,$s,t$, it only remains to check that \,$k$ \,is a morphism. 
Let \,$p\ \fleche{a}_G\ q$.\\
As \,$G$ \,is strongly connected, there exists \,$u,v \in A^*$ \,such that 
$s\ \fleche{u}_G\ p$ \,and \,$q\ \fleche{v}_G\ s$.\\
Therefore \,$uav \in {\rm L}_G(s,s) \,= \,{\rm L}_G(t,t)$. 
So there exists \,$t',t''$ \,such that 
\,$t\ \fleche{u}_G\ t'\ \fleche{a}_G\ t''$.\\
As \,$k$ \,is a bijection, we get \,$t' = k(p)$ \,and \,$t'' = k(q)$ \
hence \ $k(p)\ \fleche{a}_G\ k(q)$.
\qed\\
Another basic property for strongly connected circular or vertex-transitive 
graphs is the completeness and the co-completeness.
\enonce{{\bf Lemma}~\ref{Complete}.}
{Any strongly connected circular graph is complete and co-complete.}
Let \,$G$ \,be a strongly connected circular graph.\\
Let \,$s \in V_G$ \,and \,$a \in A_G$\,. 
There exists an edge \,$p\ \fleche{a}\ q$ \,of \,$G$.\\
As \,$G$ \,is strongly connected, there exists \,$u \in A^*$ \,such that 
\,$q\ \fleche{u}\ p$.\\
As \,$G$ \,is circular, \,$au \in {\rm L}_G(p,p) \,= \,{\rm L}_G(s,s)$.\\
So there exists a vertex \,$t$ \,such that \ $s\ \fleche{a}_G\ t$. 
Thus \,$G$ \,is complete.\\
Furthermore \,$G^{-1}$ \,remains strongly connected and circular.\\
Therefore \,$G^{-1}$ \,is complete \,{\it i.e.} \,$G$ \,is co-complete.
\qed\\[1em]
The vertex-transitivity coincides with the edge-transitivity for any
deterministic and complete graphs.
\enonce{{\bf Lemma}~\ref{EdgeTrans}.}
{For any graph \,$G$ \,deterministic and complete,\\
\hspace*{7em}$G$ \,is vertex-transitive \ \ $\Longleftrightarrow$ \ \ $G$ \,is 
edge-transitive.}
$\Longrightarrow$\,: let \,$G$ \,be a vertex-transitive deterministic graph.\\
Let us check that \,$G$ \,is edge-transitive. Let \,$s\ \fleche{a}\ t$ \,and 
\,$s'\ \fleche{a}\ t'$.\\
As \,$G$ \,is vertex-transitive, there exists an automorphism \,$h$ \,of \,$G$
\,such that \,$h(s) = s'$.\\
As \,$h$ \,is a morphism, we have \,$s' = h(s)\ \fleche{a}\ h(t)$.\\
As \,$G$ \,is deterministic, we get \,$h(t) = t'$.\\[0.25em]
$\Longleftarrow$\,: let \,$G$ \,be an edge-transitive complete graph.\\
Let us check that \,$G$ \,is vertex-transitive.
Let \,$s,s' \in V_G$ \,and \,$a \in A_G$\,.\\
As \,$G$ \,is complete, there exists \,$t,t'$ \,such that \,$s\ \fleche{a}\ t$ 
\,and \,$s'\ \fleche{a}\ t'$.\\
As \,$G$ \,is edge-transitive, there exists an automorphism \,$h$ \,of \,$G$ 
\,such that \,$h(s) = s'$.
\qed\\[1em]
The elementary circularity involves the circularity.
\enonce{{\bf Lemma}~\ref{CircEleCirc}.}
{Any elementary circular graph \,$G$ \,is circular with
\,${\rm L}_G\ \subseteq\ [\varepsilon]$\mbas{$E_G$}\,.}
Let \,$G$ \,be an elementary circular graph.\\[0.25em]
{\bf i)} Let us show that \,$G$ \,is circular.\\
It is sufficient to check that for every \,$n \geq 0$,
\begin{center}
${\rm L}_G(s,s) \,\cap \,A^{\leq n} \ = \ {\rm L}_G(t,t) \,\cap \,A^{\leq n}$ \ \
for all \,$s,t \in V_G$  
\end{center}
where \,$A^{\leq n} \,= \,\{\ u \in A^*\ |\ |u| \leq n\ \}$.
We prove this equality by induction on \,$n$.\\
$n = 0$\,: we have \,${\rm L}_G(s,s) \,\cap \,A^0 \,= \,\{\varepsilon\}$ \ for 
all \,$s \in V_G$.\\
$n \Longrightarrow n+1$\,: Let \,$s,t \in V_G$.\\
By induction hypothesis and by symmetry of \,$s$ \,with \,$t$, it is 
sufficient to check that
\begin{center}
${\rm L}_G(s,s) \,\cap \,A^{n+1} \ \subseteq \ {\rm L}_G(t,t) \,\cap \,A^{n+1}$.
\end{center}
Let \,$u \in {\rm L}_G(s,s)$ \,of length \,$|u| = n+1$.
Let us show that \,$u \in {\rm L}_G(t,t)$.\\
We distinguish two complementary cases below.\\
{\it Case 1}\,: \,$u \in {\rm E}_G(s)$.
By hypothesis, $u \in {\rm E}_G(t) \subset {\rm L}_G(t,t)$.\\
{\it Case 2}\,: \,$u \not\in {\rm E}_G(s)$.
There are \,$x,u',y \in A^*$ \,and \,$s' \in V_G$ \,such that
\begin{center}
$s\ \fleche{x}\ s'\ \fleche{u'}\ s'\ \fleche{y}\ s$ \ with \ $u = xu'y$ \ and 
\ $|u'| > 0$ \,minimal.
\end{center}
Thus \,$xy \in {\rm L}_G(s,s)$ \,with \,$|xy| < |u|$.
By induction hypothesis, \,$xy \in {\rm L}_G(t,t)$. There is \,$t'$ \,such 
that
\begin{center}
$t\ \fleche{x}\ t'\ \fleche{y}\ t$.
\end{center}
Furthermore \,$u' \in {\rm E}_G(s') = {\rm E}_G(t')$ \ hence \
$u = xu'y \in {\rm L}_G(t,t)$.\\[0.25em]
{\bf ii)} Let us show that \ $u \in {\rm L}_G\ \ \Longrightarrow\ \ u \
\fleche{*}$\mbas{$E_G$} \ $\varepsilon$.\\
By induction on \,$|u| \geq 0$. 
For \,$|u| = 0$ \,{\it i.e.} \,$u = \varepsilon$, we have 
\,$\varepsilon\ \fleche{*}$\mbas{$E_G$} \ $\varepsilon$
\,(and \,$\varepsilon \in {\rm L}_G$).\\
$|u| > 0$\,: We distinguish the two complementary cases below.\\
{\it Case 1}\,: \,$u \in {\rm E}_G$\,.
Thus \,$(u,\varepsilon) \in \overrightarrow{{\rm E}_G}$ \ hence \
$u \ \fleche{}$\mbas{$E_G$} \ $\varepsilon$.\\
{\it Case 2}\,: \,$u \not\in {\rm E}_G$\,. Let \,$s \in V_G$. 
As \,$u \in {\rm L}_G$\,, \,we have \,$s\ \fleche{u}_G\ s$.\\
As \,$u \not\in {\rm E}_G$\,, \,there are \,$x,u',y \in A_G^*$ \,and 
\,$s' \in V_G$ \,such that
\begin{center}
$s\ \fleche{x}_G\ s'\ \fleche{u'}_G\ s'\ \fleche{y}_G\ s$ \ with \ $u = xu'y$ 
\ and \ $|u'| > 0$ \,minimal.
\end{center}
Therefore \,$u' \in {\rm E}_G(s') = {\rm E}_G$ \,and 
\,$xy \in {\rm L}_G(s) = {\rm L}_G$\,.\\
As \,$|xy| < |u|$ \,and by induction hypothesis, 
\,$xy\ \fleche{*}$\mbas{$E_G$} \ $\varepsilon$.\\
As \,$u' \in {\rm E}_G$\,, \, we have \,$u'\ \fleche{}$\mbas{$E_G$} \ 
$\varepsilon$ \ hence \ $u = xu'y\ \fleche{}$\mbas{$E_G$} \ $xy\ 
\fleche{*}$\mbas{$E_G$} \ $\varepsilon$.
\qed\\[1em]
The converse of Lemma~\ref{CircEleCirc} is not verified for the following 
graph \,$G$\,:
\begin{center}
\includegraphics{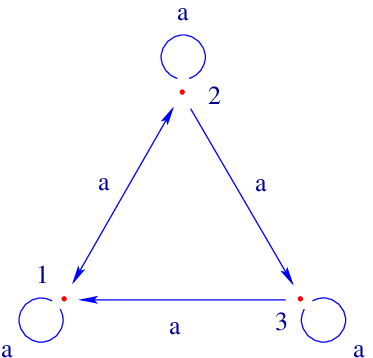}
\end{center}
since \,${\rm L}_G = a^*$ \,while 
\,${\rm E}_G(1) = {\rm E}_G(2) = \{a,aa,aaa\}$ \,and 
\,${\rm E}_G(3) = \{a,aaa\}$.\\
The circularity involves the elementary circularity for deterministic graphs.
\enonce{{\bf Lemma}~\ref{CircCircEle}.}
{Any deterministic circular graph \,$G$ \,is elementary circular with 
\,${\rm L}_G\ =\ [\varepsilon]$\mbas{$E_G$}\,.}
Let \,$G$ \,be a deterministic circular graph.\\[0.25em]
{\bf i)} Let us check that \,$G$ \,is elementary circular.\\
We have \,${\rm L}_G(s,s) = {\rm L}_G(t,t)$ \,for all \,$s,t \in V_G$.\\
Suppose that there are \,$s,t \in V_G$ \,and 
\,$u \in {\rm E}_G(s)-{\rm E}_G(t)$.\\
We have \,$u \in {\rm E}_G(s) \subset {\rm L}_G(s,s) = {\rm L}_G(t,t)$.\\
As \,$u \not\in {\rm E}_G(t)$, there are \,$x,u',y \in A^*$ \,and 
\,$t' \in V_G$ \,such that
\begin{center}
$t\ \fleche{x}\ t'\ \fleche{u'}\ t'\ \fleche{y}\ t$ \ with \ $u = xu'y$ \ and \
$|u'| > 0$.
\end{center}
As \,$u = xu'y \in {\rm L}_G(s,s)$ \,and \,$G$ \,is deterministic, there are 
single vertices \,$s',s''$ \,such that
\begin{center}
$s\ \fleche{x}\ s'\ \fleche{u'}\ s''\ \fleche{y}\ s$.
\end{center}
As \,$u' \in {\rm L}_G(t',t') = {\rm L}_G(s',s')$ \,and \,$G$ \,is 
deterministic, we get \,$s' = s''$.\\
Thus \ $u = xu'y \not\in {\rm E}_G(s)$ \ which is a contradiction.\\[0.25em]
{\bf ii)} By Lemma~\ref{CircEleCirc}, we have \
${\rm L}_G \ \subseteq \ [\varepsilon]\mbas{${\rm E}_G$} \ \subseteq \
[\varepsilon]\mbas{${\rm L}_G$}$\,.\\[0.25em]
By Lemma~\ref{ClasseVide}, ${\rm L}_G$ \,is a stable language.
By Lemma~\ref{LangStable}, 
${\rm L}_G \,= \,[\varepsilon]$\mbas{${\rm L}_G$} \,hence the equality.
\qed\\[1em]
The connectedness for finite vertex-transitive graphs implies the strong
connectedness.
\enonce{{\bf Lemma}~\ref{SymBasicFini}.}
{Any finite connected vertex-transitive graph is strongly connected.}
Let us check that \,$G$ \,is strongly connected.\\
Let \,$t\ \fleche{}\ s$. It is sufficient to show that \,$s\ \fleche{}^*\ t$.\\
As \,$G$ \,is finite, we take an elementary path 
\,$r_0\ \fleche{} \ldots \fleche{}\ r_n$ \,of maximal length \,$n$.\\
As \,$s$ \,is isomorphic to \,$r_0$\,, there is an elementary path 
\,$s = s_0\ \fleche{} \ldots \fleche{}\ s_n$\,.\\
By maximality of \,$n$, we get \,$t \in \{s_0,\ldots,s_n\}$ \,hence 
\,$s\ \fleche{}^*\ t$.
\qed\\[1em]
The finiteness of Lemma~\ref{SymBasicFini} can not be removed since the graph
\begin{center}
\begin{tabular}{ll}
 & $\{\ n+1\ \fleche{a}\ n\ |\ n \in {\mathbb Z}\ \} \,\cup 
  \,\{\ (n,\varepsilon)\ \fleche{a}\ n\ |\ n \in {\mathbb Z}\ \}$\\[0.25em]
$\cup$ & 
$\{\ (n,ui)\ \fleche{a}\ (n,u)\ |\ n \in {\mathbb Z} \,\wedge \,u \in \{0,1\}^* 
\,\wedge \,i \in \{0,1\}\ \}$
\end{tabular}
\end{center}
represented as follows:
\begin{center}
\includegraphics{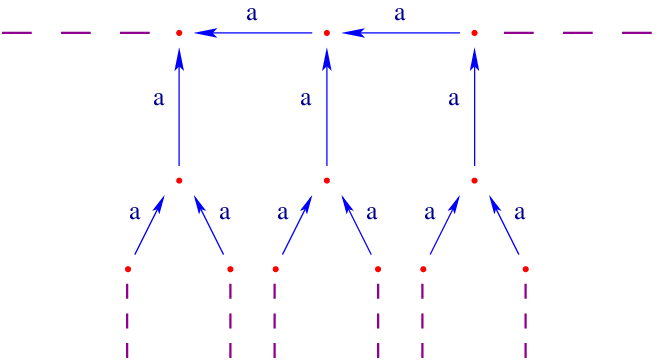}
\end{center}
is not rooted whereas it is deterministic, connected and vertex-transitive.
\\[0.25em]
Let us conclude the proof of Theorem~\ref{MainFour}
\enonce{{\bf Theorem}~\ref{MainFour}.}
{The generalized Cayley graphs are the deterministic, co-deterministic, 
vertex-transitive simple graphs.}
It remains to check that 
\,${\cal C}\inter{V_I{\croix}\mathsf{Comp},\mathsf{K}} \,\DoubleBigArrow{}_f 
\,G$.\\[0.25em]
Let us check that 
\,$G \,\subseteq \,f({\cal C}\inter{V_I{\croix}\mathsf{Comp},\mathsf{K}})$.
\\[0.25em]
Let \,$s\ \fleche{a}_G\ t$. There is a unique \,$C \in \mathsf{Comp}$ \,such
that \,$s\ \fleche{a}_C\ t$. So \,$f_C^{-1}(s)\ \fleche{a}_I\ f_C^{-1}(t)$.\\
As \,$I\ =\ {\cal C}\inter{V_I,\mathsf{H}}$\,, \,there is
\,$h \in \mathsf{H}$ \,such that \,$\inter{h} = a$ \,and 
\,$f_C^{-1}(t) \,= \,f_C^{-1}(s)\,\cdot_I\,h$. So\\[-0.75em]\mbox{}
\begin{center}
$(f_C^{-1}(s),C)\ 
\fleche{\interFootnote{(h,I)}}_{{\cal C}\interFootnote{V_I,\mathsf{K}_I}}\ 
(f_C^{-1}(t),C)$ \ \ {\it i.e.} \ \ 
$s\ \fleche{a}_{f({\cal C}\interFootnote{V_I{\croix}\mathsf{Comp},\mathsf{K}})}\ t$.
\end{center}\mbox{}\\
It remains to check that 
\,$f({\cal C}\inter{V_I{\croix}\mathsf{Comp},\mathsf{K}}) \,\subseteq \,G$.
\\[0.25em]
Let 
\,$(s,C)\ \fleche{a}_{{\cal C}\interFootnote{V_I{\croix}\mathsf{Comp},\mathsf{K}}}\ 
(t,D)$.\\[0.25em]
There is \,$h \in \mathsf{H}$ \,such that \,$\inter{h} = a$ \,and 
\,$t \,= \,s\,\cdot_I\,h$ \,and \,$C = D$.\\
So \ $s\ \fleche{\interFootnote{h}}_{{\cal C}\interFootnote{V_I,\mathsf{K}_I}}\ t$ 
\ {\it i.e.} \ $s\ \fleche{a}_I\ t$. 
Thus \,$f(s,C) \,= \,f_C(s)\ \fleche{a}_G\ f_C(t) \,= \,f(t,D)$.
\qed

\end{document}